\documentclass[aps,10pt,twocolumn,amsmath,amssymb,prb,reprint,citeautoscript,superscriptaddress,showkeys,showpacs]{revtex4-1}

\usepackage{xr}
\usepackage{amsmath}
\usepackage{amsfonts, amssymb,amsxtra}
\usepackage[]{graphicx}
\usepackage{amsfonts}
\usepackage{xcolor}

\begin{document}
\title{Post-quench gap dynamics of two-band superconductors}
\author{Tianbai Cui}
\affiliation{School of Physics and Astronomy, University of Minnesota, Minneapolis,
MN 55455, USA}
\author{Michael Sch\"utt}
\affiliation{Condensed Matter Theory Group, Paul Scherrer Institute, CH-5232 Villigen
PSI, Switzerland}
\author{Peter P. Orth}
\affiliation{Department of Physics and Astronomy, Iowa State University, Ames,
Iowa 50011, USA}
\affiliation{Ames Laboratory, Ames, Iowa 50011, USA}
\author{Rafael M. Fernandes}
\affiliation{School of Physics and Astronomy, University of Minnesota, Minneapolis,
MN 55455, USA}
\begin{abstract}
Recent experimental progress in the fields of cold quantum gases and
ultrafast optical spectroscopy of quantum materials allows to controllably
induce and probe non-adiabatic dynamics of superconductors and superfluids.
The time-evolution of the gap function before relaxation with the
lattice is determined by the superposition of coherently evolving
individual Cooper pairs within the manifold of the Bardeen-Cooper-Schrieffer
(BCS) wavefunction. While dynamics following an abrupt quench of the
pairing interaction strength in the single-band BCS model has been
exactly solved due to the integrability of the model, the dynamics
of post-quench multi-band superconductors remain under scrutiny. Here,
we develop a generalization of the Volkov-Kogan Laplace-space perturbative
method that allows us to determine the non-adiabatic gap dynamics
of two-band fully gapped superconductors for a wide range of quench
amplitudes. Our approach expands the long-time dynamics around the
steady-state asymptotic value of the gap, which is self-consistently
determined, rather than around the equilibrium value of the gap. We
explicitly demonstrate that this method recovers the exact solution
of the long-time gap dynamics in the single-band case and perfectly
agrees with a numerical solution of the two-band model. We discover
that dephasing of Cooper pairs from different bands leads to faster
collisionless relaxation of the gap oscillation with a power-law of
$t^{-3/2}$ instead of the well-known $t^{-1/2}$ behavior found in
the single-band case. Furthermore, the gap oscillations display beating
patterns arising from the existence of two different asymptotic gap
values. Our results have important implications to a variety of two-band
superconductors driven out of equilibrium, such as iron-based superconductors,
MgB$_{2}$, and SrTiO$_{3}$.
\end{abstract}
\maketitle

\section{Introduction}

\label{sec:introduction} Superconductors that are perturbed into
a state away from equilibrium display an extremely rich and interesting
dynamical behavior. This originates from the interplay between the
dynamics of its fermionic quasi-particle excitations and that of the
superconducting order parameter, as expressed, for example, in the
superconducting gap equation. Close to equilibrium, various collective
modes emerge such as the Anderson-Bogoliubov phase mode~\cite{andersonRandomPhaseApproximationTheory1958}
and the longitudinal Schmid (or Higgs) amplitude mode~\cite{schmidTimeDependentGinzburgLandau1966,pekkerAmplitudeHiggsModes2015},
which describe phase and amplitude fluctuations of the order parameter.
The transverse Carlson-Goldman mode describes the coupled oscillations
of normal currents and supercurrents~\cite{carlsonPropagatingOrderParameterCollective1975,Pals-PhysRep-1989},
whereas in multi-gap superconductors additional Leggett phase modes
appear~\cite{leggettNumberPhaseFluctuationsTwoBand1966}, corresponding
to oscillations of the relative phases of the different gaps. Interesting
dynamics also occurs farther away from equilibrium, where one observes,
for example, intriguing non-linear behaviors such as dynamic instabilities
towards slowly damped~\cite{Volkov1974,Galperin_Kozub_Spivak-JETP-1981}
or even undamped order parameter oscillations~\cite{Barankov-Rabi-Oscillations,Yuzbashyan_Altshuler_Enolskii-PRB-2005,Yuzbashyan-2006-PRL,Barankov-Synchronization}.

Generally, the dynamic response of a superconductor depends on the
type of perturbation that is applied, for example, whether it is adiabatic
or non-adiabatic, linear or non-linear, and whether it is charge neutral
or charged. It also depends on the hierarchy of a number of important
timescales such as the quasi-particle energy relaxation time $\tau_{\varepsilon}$,
the dynamical scale of the superconducting order parameter $\tau_{\Delta}$,
the timescale of the external perturbation $\tau_{\text{pert}}$ and
the characteristic observation time $t$~\cite{aronovBoltzmannequationDescriptionTransport1981,langenbergNonequilibriumSuperconductivity1986,kopninTheoryNonequilibriumSuperconductivity2001}.
Here, we are interested in the case of $\tau_{\Delta}\ll\tau_{\varepsilon}$
and in fast, non-adiabatic perturbation occurring on a timescale $\tau_{\text{pert}}\ll\tau_{\Delta}\approx t$.
This non-adiabatic, collisionless regime has been explored in a linearized
approach close to equilibrium in the seminal work by Volkov and Kogan~\cite{Volkov1974},
who studied the gap dynamics of a single-band superconductor following
a small and instantaneous perturbation. They found coherent gap oscillations
that are only algebraically damped $\propto t^{-1/2}$, analogous
to Landau damping in a collisionless plasma~\cite{landauVibrationsElectronicPlasma1946,Kamenev-NonEqFieldTheory-Book}.
More recently, experimental progress on two distinct fronts have brought
renewed interest to this field: (1) Ultrafast optical studies in the
Terahertz regime have unveiled non-adiabatic, coherent gap dynamics
in thin superconducting films~\cite{shimanoHiggsModeSuperconductors2019},
for example, in NbN~\cite{Shimano-PRL-2012,Shimano-PRL-2013,Shimano-Science-2014,PhysRevLett.107.177007}
and Nb$_{3}$Sn~\cite{Yang-NatureMaterials-2018,Cui2018,Yang-NaturePhotonics-2019};
(2) Cold-atom realizations of superfluids and Bose-Einstein condensates
have provided a fruitful avenue to induce non-adiabatic dynamics by
performing rapid parameter changes such as quenching the pairing interaction
strength~\cite{langenUltracoldAtomsOut2015,bloch:885}.

The situation of a rapid parameter quench is theoretically particularly
interesting as it is amenable to analytical approaches. Going beyond
the linear analysis of Volkov and Kogan and exploiting the integrability
of the Bardeen-Cooper-Schrieffer (BCS) Hamiltonian~\cite{Richardson-NuclPhys-1964,gaudinBetheWavefunction2014,Dukelsky-RMP-2004},
a number of works have explored post-quench non-adiabatic dynamics
of single-band BCS superconductors far-away from equilibrium~\cite{Barankov-Rabi-Oscillations,Yuzbashyan_Altshuler_Enolskii-PRB-2005,Yuzbashyan-2005-JPA,Barankov-Synchronization,Yuzbashyan_Dzero_Gurarie_Foster-PRA-2015,Yuzbashyan_Altshuler-PRB-2005}.
It was discovered that non-equilibrium dynamics at times $\tau_{\Delta}\ll t\ll\tau_{\varepsilon}$
fall into one of three distinct classes (or "phases")~\cite{Barankov-Synchronization,YuzbashyanDzero-PRL-2006},
which can be topologically distinguished by the number of complex
roots of the spectral polynomial~\cite{Yuzbashyan_Altshuler_Enolskii-PRB-2005,Yuzbashyan-2006-PRL}:
Phase I, where the gap decays exponentially to zero; Phase II, where
the gap oscillates with frequency $2\Delta_{\infty}$ and decays algebraically
$\propto t^{-1/2}$ to a finite value $\Delta_{\infty}$; and phase
III, where persistent undamped gap oscillations occur. Phase II in
the non-equilibrium quench phase diagram~\cite{Barankov-Synchronization,YuzbashyanDzero-PRL-2006}
contains the linear regime around equilibrium studied by Volkov and
Kogan~\cite{Volkov1974}. Finally, we note that the topological classification
explains why terahertz induced gap dynamics is qualitatively similar
to the case of a parameter quench, as has been observed in various
numerical studies~\cite{Chou_Foster-PRB-2017,Papenkort-PRB-2007,Papenkort_JPhys-2009,krullSignaturesNonadiabaticBCS2014a,Eremin-2band,Cui2018}.

In this paper, we extend these previous studies by addressing numerically
and analytically the gap dynamics of two-band superconductors following
an interaction quench. Our motivation is on the fact that multi-band
superconductivity is realized in a variety of materials with conventional
and unconventional pairing mechanisms. Primary examples are MgB$_{2}$~\cite{budkoSuperconductivityMagnesiumDiboride2015},
the iron-based superconductors~\cite{Chubukov-PRB-2008}, Sr$_{2}$RuO$_{4}$~\cite{mackenzieSuperconductivityMathrmSrMathrmRuO2003},
heavy fermions~\cite{wirthExploringHeavyFermions2016}, strontium
titanate \cite{Fernandes13,Trevisan18}, and oxide heterostructures
such as LaAlO$_{3}$/SrTiO$_{3}$~\cite{Scheurer-NatComm-2015}.
Unconventional multi-orbital superfluidity has also been reported
in cold-atom setups on the honeycomb lattice~\cite{soltan-panahiQuantumPhaseTransition2012a}.
While different superconducting gap symmetries are possible in the
presence of multiple Fermi surfaces, we will focus on the simplest
case of $s$-wave superconductivity. As the quench dynamics is identical
for $s^{+-}$ and $s^{++}$ pairing, corresponding to gaps with opposite
or same signs on the two Fermi surfaces, respectively, our results
apply to both cases. Quenches in two-band $s$-wave superconductors
have so far only been studied numerically~\cite{Eremin-2band,Krull2016-Leggett-Higgs,Dzero-PRB-2015},
focusing on the coupling between the Higgs and the Leggett mode~\cite{Krull2016-Leggett-Higgs}
or the competition between superconductivity and spin-density wave
order~\cite{Dzero-PRB-2015}. Generalizations to quenches between
other pairing symmetries such as time-reversal symmetry breaking $s+is$
or $s+id$ pairing are interesting avenues for further work. Indeed,
a recent numerical study of Terahertz induced gap dynamics for $s+is$
pairing has revealed an unusual coupling between the Higgs amplitude
and the Leggett (relative) phase mode~\cite{Mueller-PRB-2018}.

\begin{figure}[tbh]
\begin{centering}
\includegraphics[width=1\columnwidth]{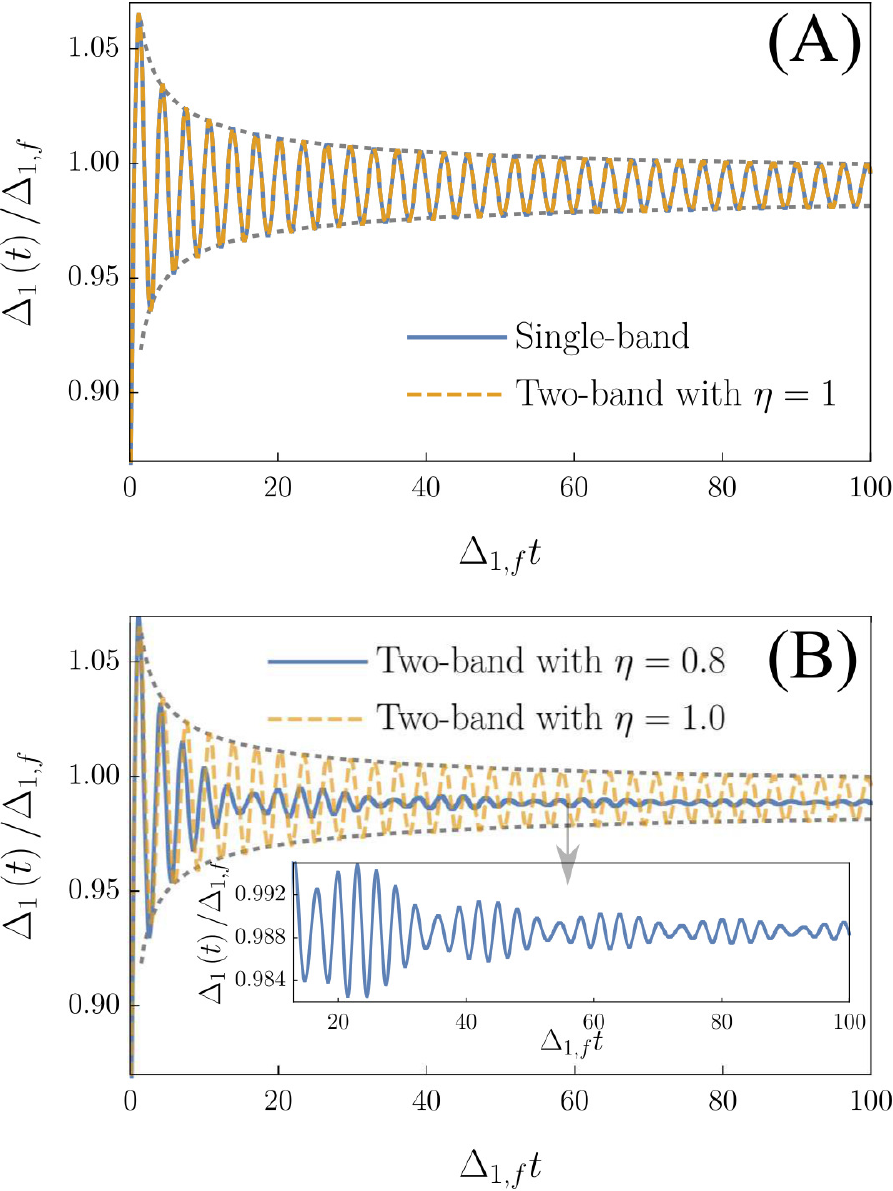} 
\par\end{centering}
\caption{Summary of our main results for the gap dynamics of a quenched two-band
superconductor. In these figures, only inter-band pairing is included.
(panel A) When the densities of states of the two bands are the same,
$\eta\equiv\frac{\mathcal{N}_{1}}{\mathcal{N}_{2}}=1$, the behavior
is the same as that of a single-band model. (panel B) When $\eta\protect\neq1$,
the behavior is different in that the damping of the gap oscillations
changes from $t^{-1/2}$ to $t^{-3/2}$ and a beating pattern occurs
due to the existence of two oscillation frequencies (inset). In this
figure, $\Delta_{1}$ is the gap of band $1$ and $\Delta_{1f}$ is
the quenched value of the gap. The parameters used here were $v_{i}=0.19$,
$v_{f}=0.2$ for both panel A and B.\label{fig:summary}}
\end{figure}

Exact solutions of the time-dependent two-band BCS model only exist
for special fine tuned values of the intra- and inter-band interaction
parameters, where the problem effectively reduces to the single-band
case (see below and Ref.~\onlinecite{Dzero-PRB-2015}). It is an
open question whether the generic two-band BCS model is integrable.
Here, we develop a generalization of the Volkov-Kogan Laplace-space
analysis in order to investigate the non-adiabatic post-quench dynamics
in generic two-band BCS models. Like Volkov and Kogan we solve linearized
equations of motion in Laplace space, but an important distinction
of our work is that we expand around the long-time steady-state of
the system instead of the equilibrium state. This allows us to explore
the gap dynamics away from the weak-quench limit in a larger region
of the non-equilibrium phase diagram. We achieve this methodological
advancement by self-consistently solving for the steady-state value
of the superconducting gap $\Delta_{\infty}$. We show in detail that
our method reproduces the exact solution in phase II of the single-band
model. For the two-band model we carefully check our analytical results
by comparing to the numerical solution of the dynamics. We find that
the oscillatory gap dynamics exhibits pronounced beating behavior
due to the presence of two asymptotic gap values $\Delta_{1,\infty}$
and $\Delta_{2,\infty}$, which has been previously reported in a
numerical investigation of Terahertz driven gap oscillations in two-band
superconductors~\cite{Eremin-2band}. A central new result of our
work is that the decay of the gap oscillations due to Landau damping
in two-band superconductors is governed by a power-law $\propto t^{-3/2}$
that is different from the one found in the single band case, where
it is $\propto t^{-1/2}$ (see Fig. \ref{fig:summary}). Earlier numerical
studies of multi-gap superconductors have reported power-law decays
of $t^{-1/2}$, although in that case the dynamics was driven not
a by an interaction quench, but by laser pulses \cite{Eremin-2band}.
Interestingly, faster than $t^{-1/2}$ decay was also seen in the
case of superconducting nanowires, where electronic subbands arise
due to confinement ~\cite{Zachmann_NJPhys-2013}. Finally, a similar
$t^{-3/2}$ decay of the pairing amplitude has been found in quenches
into the strong pairing (Bose-Einstein condensation (BEC)) regime
in three dimensions, but by a different microscopic mechanism~\cite{Gurarie-BEC-t-three-halves,Yuzbashyan_Dzero_Gurarie_Foster-PRA-2015}.

The remainder of the paper is organized as follows: in Sec.~\ref{sec:model},
we define the two-band BCS model and formulate it in terms of Anderson
pseudospins. We then derive equations of motion of the pseudospins
that govern the non-adiabatic dynamics of individual Cooper pairs
and the gap following an instantaneous quench of the BCS coupling
strength. In Sec.~\ref{sec:numerical_results}, we present numerical
solutions of the gap dynamics in the regime of weak quenches, which
show the main features of oscillatory beating and algebraic decay
$\propto t^{-3/2}$. In Sec.~\ref{sec:long_time_gap_dynamics}, we
present our main analytical calculation and flesh out the details
of our method to find the long-time dynamics of the gap using a self-consistent
Laplace analysis. In Sec.~\ref{subsec:linearized_eom}, we derive
linearized equations of motion around the long-time steady-state.
We present the solution of these equations in Laplace space in Sec.~\ref{subsec:sol_laplace},
which depends on the steady-state values of the gap $\Delta_{\alpha,\infty}$
and the pseudo-spins $S_{\alpha,\infty}^{i}$. These values are determined
in Sec.~\ref{subsec:gap_from_self_consistency} by solving self-consistent
equations via an ansatz for the non-equilibrium distribution function
in the steady-state. We first show that our method yields the exact
solution in the single-band model, and then apply it to the two-band
case, where only numerical solutions are available. Finally, in Sec.~\ref{subsec:gap_oscillations},
we discuss the long-time gap dynamics in real-time by performing an
inverse Laplace transformation. We explicitly show how the new power-law
decay exponent emerges from a distinct analytical structure of the
gap in Laplace space and demonstrate how one re-obtains the single-band
result. We conclude in Sec.~\ref{sec:conclusions}, and present additional
details of our analytical calculations in the Appendices.

\section{BCS model and quench protocol}

\label{sec:model}

\subsection{Pseudospin formalism for equilibrium two-band superconductors}

\label{subsec:pseudospin_formalism} 
\begin{figure}[tbh]
\begin{centering}
\includegraphics[width=1\columnwidth]{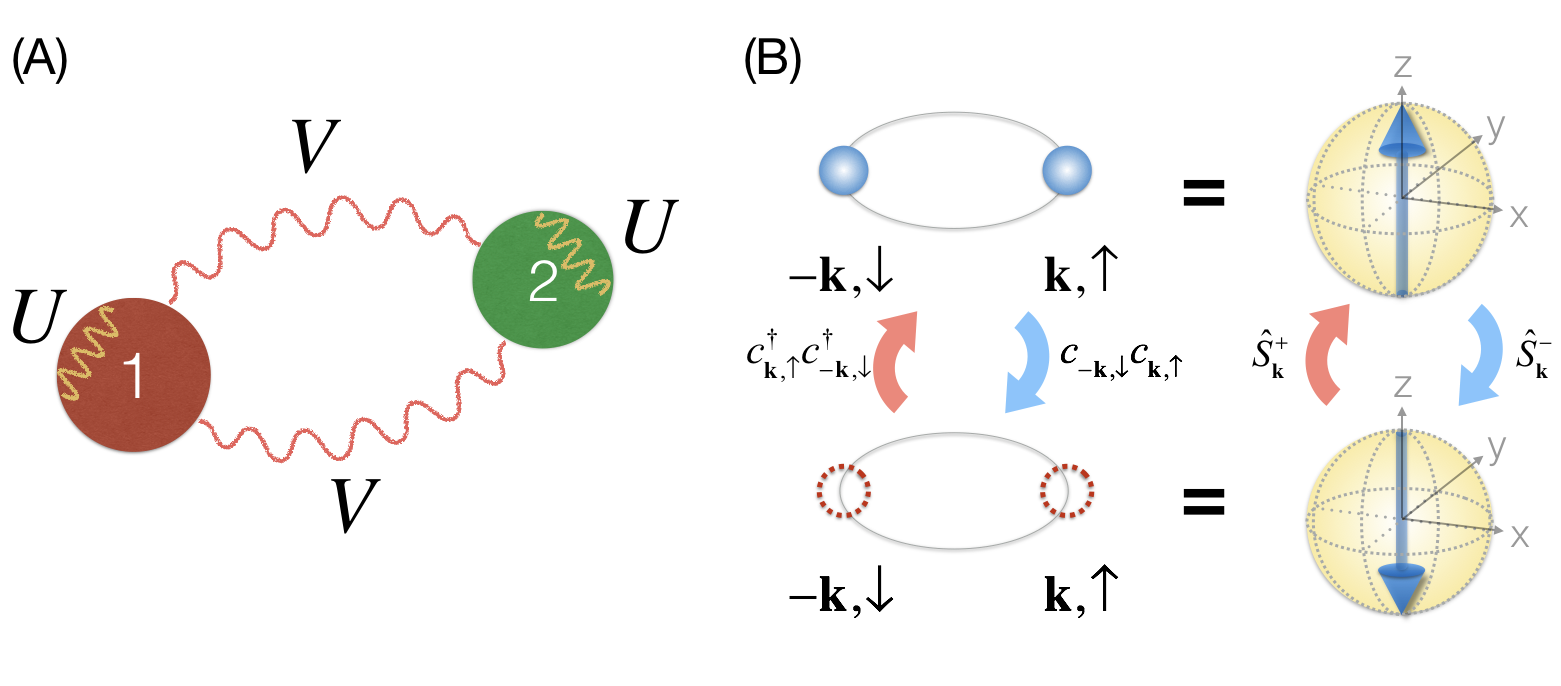} 
\par\end{centering}
\caption{(A) Schematics of the two bands and the interactions between them.
(B) Schematics of the mapping between the electronic operators and
the pseudo-spin operators.}
\label{Fig_FS_PS} 
\end{figure}

We start from the reduced BCS
Hamiltonian~\cite{BCS} for two-band superconductors 
\begin{eqnarray}
H_{\text{BCS}} & = & \sum_{\mathbf{k},\sigma,\alpha}\varepsilon_{\mathbf{k},\alpha}c_{\mathbf{k},\sigma,\alpha}^{\dagger}c_{\mathbf{k},\sigma,\alpha}\nonumber \\
 &  & +\frac{1}{N}\sum_{\mathbf{k},\mathbf{p},\alpha,\beta}V_{\alpha\beta}c_{\mathbf{k},\uparrow,\alpha}^{\dagger}c_{-\mathbf{k},\downarrow,\alpha}^{\dagger}c_{-\mathbf{p},\downarrow,\beta}c_{\mathbf{p},\uparrow,\beta}
\end{eqnarray}
where $\alpha,\,\beta\in\left\{ 1,\,2\right\} $ are the band indices,
$\varepsilon_{\mathbf{k},\alpha}$ is the electronic dispersion near
the Fermi level in band $\alpha$ (including the chemical potential),
and $V_{\alpha\beta}$ is the effective pairing interaction between
band $\alpha$ and band $\beta$. Although not important in the following,
one may assume parabolic dispersions, $\varepsilon_{\mathbf{k},\alpha}=\mathbf{k}^{2}/2m_{\alpha}-\mu$.
The interaction constants $V_{\alpha\beta}$ are positive (negative)
if the interaction is repulsive (attractive). In multi-band systems,
different bands develop different values of the superconducting gap,
depending on the values of the intra-band interactions, $V_{11}$
and $V_{22}$, and the inter-band interactions, $V_{12}$ and $V_{21}$
(see Fig. \ref{Fig_FS_PS} (A)) as well as the density of states of
the two bands at the Fermi level, $\mathcal{N}_{\alpha}$. We assume
that the two bands have the same intra-band electronic interactions
such that $V_{11}=V_{22}\equiv U$; by definition, $V_{12}=V_{21}\equiv V$.
Due to the different density of states $\mathcal{N}_{1}\neq\mathcal{N}_{2}$,
electrons in different bands experience different effective interaction
strengths. The BCS gap equation is therefore band-dependent: 
\begin{equation}
\Delta_{\alpha}=\Delta_{\alpha}'+i\Delta_{\alpha}''=-\frac{1}{N}\sum_{\mathbf{p},\beta}V_{\alpha\beta}\left\langle c_{-\mathbf{p},\downarrow,\beta}c_{\mathbf{p},\uparrow,\beta}\right\rangle 
\end{equation}
Going from summation over momenta to integrations over energy using
the density of states, we write the equilibrium BCS gap equations
explicitly in matrix form in the band-space. 
\begin{equation}
\left(\begin{array}{c}
\Delta_{1}\\
\Delta_{2}
\end{array}\right)=\hat{\gamma}v\left(\begin{array}{c}
\int_{-\Lambda}^{\Lambda}d\varepsilon\frac{\Delta_{1}}{2E_{1}}\tanh\left(\frac{E_{1}}{2T}\right)\\
\int_{-\Lambda}^{\Lambda}d\varepsilon\frac{\Delta_{2}}{2E_{2}}\tanh\left(\frac{E_{2}}{2T}\right)
\end{array}\right)
\end{equation}
where $\Lambda$ is a high-energy cutoff and 
\begin{equation}
\hat{\gamma}=\left(\begin{array}{cc}
r & -\eta\\
-1 & r\eta
\end{array}\right)\label{eq:gamma_matrix}
\end{equation}
with $\eta=\mathcal{N}_{2}/\mathcal{N}_{1}$ being the ratio of the
density of states of the two bands, $E_{\alpha}=\sqrt{\varepsilon^{2}+\Delta_{\alpha}^{2}}$
is the Bogoliubov quasiparticle dispersion in band $\alpha$ and $T$
is the temperature of the system. In the following, we restrict our
analysis to the $T=0$ ground state as the initial pre-quench state
of the system. We have also defined the dimensionless inter-band interaction
coupling constant $v=V\mathcal{N}_{1}$, and the dimensionless ratio
$r=-U/V$ between intra-band and inter-band interactions. Here, we
include the minus sign in the definition, as we will assume that $U<0$
is negative, corresponding to attractive intra-band interaction.

Note that the ratio of the density of states in the two bands, $\eta=\mathcal{N}_{2}/\mathcal{N}_{1}$,
determines the relative sizes of the superconducting gaps of the two
bands. If the two bands have the same density of states near the Fermi
energy, i.e. $\eta=1$, the matrix $\hat{\gamma}$ becomes symmetric.
Therefore, the gap equations are solved by $\Delta_{1}=-\Delta_{2}$
for repulsive inter-band interaction ($v>0$), corresponding to $s^{+-}$
pairing, and $\Delta_{1}=\Delta_{2}$ for attractive inter-band interaction
($v<0$), corresponding to $s^{++}$ pairing. In this paper, we will
focus on the case with $\eta\neq1$, in which case the amplitude of
the two gaps is different in equilibrium $|\Delta_{1}|\neq|\Delta_{2}|$
and the multi-band nature of the system has a pronounced imprint on
the non-equilibrium dynamics of the superconducting gap.

It is convenient to use the pseudospin formalism~\cite{Anderson-RPA-BCS}
to study the non-equilibrium dynamics of the superconducting state.
In the mean-field approach, which is exact in the BCS regime we consider
here, the BCS Hamiltonian can be described by pseudospins exposed
to an effective magnetic field: 
\begin{equation}
H_{\text{BCS}}=-\sum_{\mathbf{k},\alpha}\mathbf{B}_{\mathbf{k},\alpha}\cdot\hat{\mathbf{S}}_{\mathbf{k},\alpha}+\text{const.}\label{PS_BCS}
\end{equation}
where $\mathbf{B}_{\mathbf{k},\alpha}=2\left(\Delta_{\alpha}^{'},-\Delta_{\alpha}^{''},-\varepsilon_{\mathbf{k},\alpha}\right)$\textcolor{blue}{{}
}and 
\begin{align}
\hat{S}_{\mathbf{k},\alpha}^{-} & =c_{-\mathbf{k},\downarrow,\alpha}c_{\mathbf{k},\uparrow,\alpha}\\
\hat{S}_{\mathbf{k},\alpha}^{+} & =c_{\mathbf{k},\uparrow,\alpha}^{\dagger}c_{-\mathbf{k},\downarrow,\alpha}^{\dagger}\\
\hat{S}_{\mathbf{k},\alpha}^{z} & =\frac{1}{2}\left(c_{\mathbf{k},\uparrow,\alpha}^{\dagger}c_{\mathbf{k},\uparrow,\alpha}+c_{-\mathbf{k},\downarrow,\alpha}^{\dagger}c_{-\mathbf{k},\downarrow,\alpha}-1\right)
\end{align}
The constant term contributes to the condensation energy, which will
be ignored because it is not relevant to the dynamics out of equilibrium.
The mapping between pseudo-spins and electronic pair operators is
summarized in Fig. \ref{Fig_FS_PS}(B). The anti-commutation relation
between the electronic operators ensures the spin commutation relation
between $\mathbf{\hat{S}}_{\mathbf{k},\alpha}$. Notice that despite
the simple form of the pseudospin Hamiltonian, the effective magnetic
field is self-consistently determined by the pseudospins collectively
via the gap equation: 
\begin{equation}
\Delta_{\alpha}=-\frac{1}{N}\sum_{\mathbf{k},\beta}V_{\alpha\beta}S_{\mathbf{k},\beta}^{-}\label{gap_eq}
\end{equation}
where $S_{\mathbf{k},\alpha}^{-}=\left\langle \hat{S}_{\mathbf{k},\alpha}^{-}\right\rangle =\left\langle c_{-\mathbf{k},\downarrow,\alpha}c_{\mathbf{k},\uparrow,\alpha}\right\rangle $.

In equilibrium, the pseudospins are parallel to the effective magnetic
field. It is convenient to work in a gauge where both the gaps are
real. Then the expectation values of the pseudo-spins at temperature
$T$ are given by \begin{subequations} 
\begin{align}
S_{\mathbf{k},\alpha}^{x} & =\frac{\Delta_{\alpha}}{2E_{\alpha}}\tanh\left(\frac{E_{\alpha}}{2T}\right)\\
S_{\mathbf{k},\alpha}^{y} & =0\\
S_{\mathbf{k},\alpha}^{z} & =\frac{-\varepsilon_{\mathbf{k}}}{2E_{\alpha}}\tanh\left(\frac{E_{\alpha}}{2T}\right)\,.\label{aux}
\end{align}
\end{subequations} Note that the length of the pseudospins in equilibrium
is determined by the Fermi-Dirac distribution, $n_{\text{F}}$, of
the Bogoliubov quasiparticles, i.e. $\left|\mathbf{S}_{\mathbf{k},\alpha}\right|=\frac{1}{2}-n_{\text{F}}$.
As mentioned above, we will focus hereafter on initial pre-quench
states at zero temperature ($T=0$).

\subsection{Equations of motion for the pseudospins}

\label{subsec:eom} We consider the situation where the system is
driven out of equilibrium by a sudden quench of the pairing interaction.
Specifically, we focus on a sudden change of the inter-band coupling
$v_{i}\rightarrow v_{f}$ while keeping the ratios between intra-
and inter-band interactions, $r=U/V$, and between the densities of
states, $\eta$, unchanged, i.e., $r_{i}=r_{f}$ and $\eta_{i}=\eta_{f}$.
The subscript $i$ and $f$ denote the initial and final values of
the respective dimensionless constants. Note that this requires quenching
both intra- and inter-band interactions $U$ and $V$ in such a way
to keep their ratio $r$ fixed. We focus on these quench protocols
to constrain the parameter space. Generally, one can also consider
quenches of $r$, however, this is expected to not lead to qualitative
changes to the non-equilibrium dynamics, as it corresponds to a different
way to prepare the initial conditions.

If the two bands have different densities of states, i.e. $\eta\neq1$,
the quench dynamics is intrinsically different from single-band systems.
In the pseudospin formalism, the superconducting gap determines the
intrinsic frequency of the pseudospin precession. Therefore, once
the two bands have different densities of states, they develop different
values of the gap, leading to two distinct intrinsic frequencies.
In addition, the gap also serves as the effective magnetic field that
drives the precession. Through the inter-band interaction, each band
experiences an oscillating magnetic field with the intrinsic frequency
of the other band. Hence, the dephasing of the pseudospin oscillations
in multi-band systems is fundamentally different from single-band
systems. The dynamics is described by two sets of equations of motion
for the two bands, which are derived from Eq. (\ref{PS_BCS}) in terms
of expectation values of the pseudospins operators, 
\begin{equation}
\frac{d}{dt}\mathbf{S}_{\mathbf{k},\alpha}\left(t\right)=\mathbf{S}_{\mathbf{k},\alpha}\left(t\right)\times\mathbf{B}_{\mathbf{k},\alpha}\left(t\right)\label{EOM}
\end{equation}
which are similar to the one-band case, but now with an extra band
index $\alpha$. More importantly, the pseudospin dynamics in the
two bands are coupled via the gap equations with a time-dependent
inter-band coupling strength $v\left(t\right)=v_{i}\theta\left(-t\right)+v_{f}\theta\left(t\right)$:

\begin{figure*}
\begin{centering}
\includegraphics[width=0.8\paperwidth]{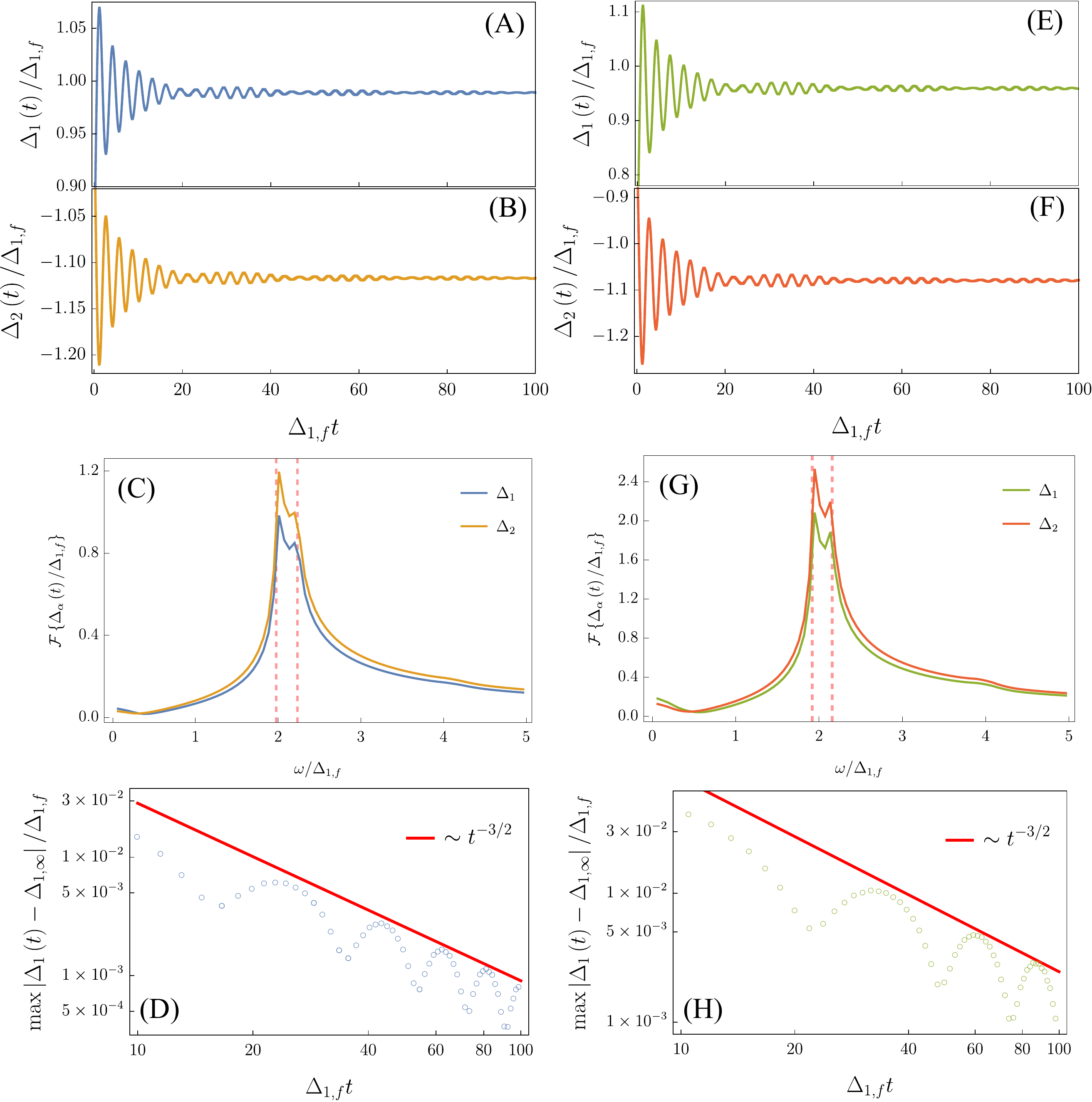}
\par\end{centering}
\caption{Numerical results for the gap oscillations in two-band superconductors.
(A)-(D) are the results for an interaction quench from $v_{i}=0.19$
to $v_{f}=0.2$. (E)-(H) correspond to aninteraction quench from $v_{i}=0.18$
to $v_{f}=0.2$. (C) and (G) are the Fourier spectrums of the gap
oscillations. (D) and (H) shown the $t^{-3/2}$ damping of the gap
oscillations in a log-log plot. The ratio of the density of states
between the two bands is $\eta=0.8$ in these calculations.}
\label{Fig_Gap_Oscillations}
\end{figure*}
\begin{equation}
\Delta_{\alpha}\left(t\right)=v\left(t\right)\sum_{\beta}\gamma_{\alpha\beta}\int d\varepsilon S_{\beta}^{-}\left(\varepsilon,t\right)\label{gap_eq_energy_space}
\end{equation}
The equations of motion for the pseudospins, Eq. (\ref{EOM}) and
the time-dependent gap equation, Eq. (\ref{gap_eq_energy_space}),
determine the post-quench gap dynamics of two-band superconductors.

In the following, we first solve these equations numerically and describe
our results. Then, we analytically find the long-term asymptotic behavior
of the gap oscillations using Laplace transforms. We develop a generalization
of the well-known procedure pioneered by Volkov and Kogan in Ref.~\onlinecite{Volkov1974}
(see also Refs. \cite{YuzbashyanDzero-PRL-2006,Yuzbashyan_Dzero_Gurarie_Foster-PRA-2015}).
By expanding around the long-time \emph{non-equilibrium} pseudo-spin
steady state, instead of the final equilibrium state, we are able
to not only determine the power-law decay of the gap oscillations,
but also the steady-state non-equilibrium gap values $\Delta_{\alpha,\infty}$.
We also explicitly show how our solution approaches the known single-band
result as $\eta\rightarrow1$.

\section{Numerical results for the post-quench gap dynamics}

\label{sec:numerical_results} We solve the equations of motion~\eqref{EOM},
together with the gap equation~\eqref{gap_eq_energy_space}, numerically
using the Runge-Kutta method. We focus on the weak-quench limit, to
later compare with our analytical expansion. Results for two different
ratios of initial and final inter-gap couplings $v_{i}/v_{f}=0.95$
and $0.9$ (with fixed $v_{f}=0.2$) are shown in Fig. \ref{Fig_Gap_Oscillations}.
The other parameters are kept fixed: $r_{i}=r_{f}=0$, $\eta=0.8$,
$T_{i}=0$. In equilibrium, this corresponds to the following gap
ratios $\Delta_{1,i}/\Delta_{2,i}=-0.8852$ for $v_{i}=0.19$ , $\Delta_{1,i}/\Delta_{2,i}=-0.8857$
for $v_{i}=0.18$ and $\Delta_{1,f}/\Delta_{2,f}=-0.8847$ for $v_{f}=0.2$.
The figure contains both the time traces of the gap oscillations as
well as their Fourier transforms.

There are two important qualitative features that emerge in the two-band
case: first, the gap oscillations are characterized by two frequencies,
corresponding to the steady-state values $\Delta_{1,\infty}$ and
$\Delta_{2,\infty}$. This leads to pronounced beating when these
two frequencies are sufficiently close to each other. This phenomenon
has been described previously in numerical studies of two-band (multi-band)
superconductors exposed to terahertz laser pulses~\cite{Eremin-2band,Krull2016-Leggett-Higgs,Zachmann_NJPhys-2013}
Second, the algebraic decay of the gap oscillations ($\propto t^{-\alpha}$)
occurs more rapidly than in the single-band case. We numerically determine
the exponent to be $\alpha_{\text{2-band}}=3/2$ as opposed to $\alpha_{\text{1-band}}=1/2$.

\begin{figure}[tbh]
\begin{centering}
\includegraphics[width=1\columnwidth]{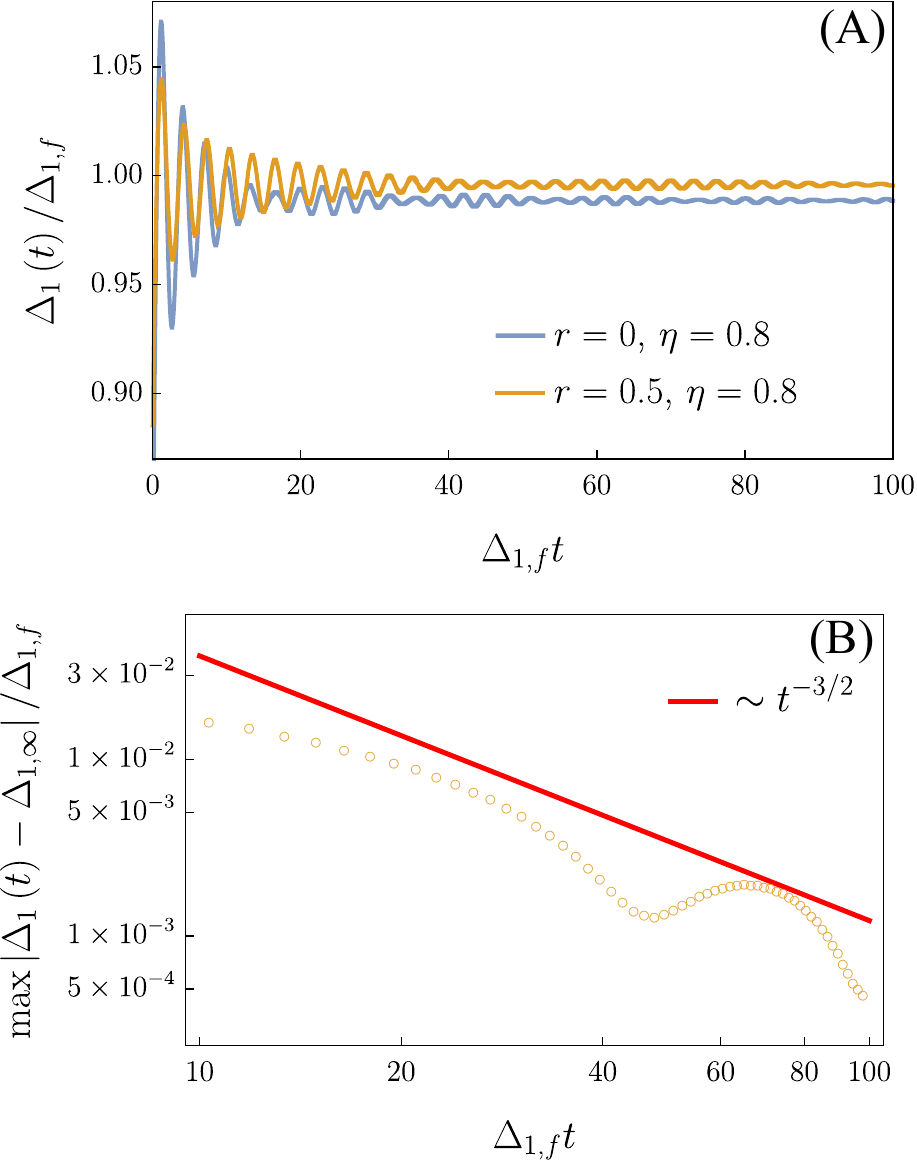} 
\par\end{centering}
\caption{(A) Gap oscillations for the case of inter-band pairing only ($r=0$)
and inter-band and intra-band pairing ($r=0.5$). Here, we set $\eta=0.8$.
(B) The $t^{-3/2}$ damping of the gap oscillations in a log-log scale
for the case $r=0.5$.}
\label{Fig_Gap_Oscillations_r} 
\end{figure}

This behavior seems insensitive to the actual value of $r$. In Fig.
\ref{Fig_Gap_Oscillations_r}, we compare the behavior of $\Delta_{1}(t)$
for the cases in which $r=0.5$ and $r=0$. The other parameters used
were $v_{i}=0.19$ and $v_{f}=0.2$\textcolor{black}{.} We note that
an exponent of $\alpha=3/2$ also emerges if one considers deep quenches
into the Bose-Einstein condensate (BEC) regime in a three-dimensional
system~\cite{Gurarie-BEC-t-three-halves,Yuzbashyan_Dzero_Gurarie_Foster-PRA-2015}.

\section{Long-time asymptotic gap dynamics}

\label{sec:long_time_gap_dynamics} In order to gain more insights
on the transient dynamics of the superconducting gap in two-band systems,
it is instructive to have analytic solutions for the superconducting
gap evolution. The gap dynamics in single-band conventional superconductors
with isotropic gap structures can be solved exactly due to the integrability
of the BCS model~\cite{Richardson-NuclPhys-1964,Dukelsky-RMP-2004,Yuzbashyan-2005-JPA,Yuzbashyan_Altshuler-PRB-2005,Yuzbashyan_Altshuler_Enolskii-PRB-2005,Yuzbashyan-2006-PRL,Barankov-Synchronization,Yuzbashyan_Dzero_Gurarie_Foster-PRA-2015}.
The two-band BCS model doubles the number of degrees of freedom compared
to the single-band model. Due to the coupling between the two distinct
bands, the integrals of motion that were constructed previously for
the single-band BCS model~\cite{Yuzbashyan_Altshuler-PRB-2005,Yuzbashyan_Dzero_Gurarie_Foster-PRA-2015}
do not commute between the two bands, except in the symmetric case
$\eta=1$. In the single-band case, it was determined that there are
three different ``phases'' depending on the strength of the quench
$\Delta_{i}/\Delta_{f}$: in phase I, corresponding to $\Delta_{i}/\Delta_{f}>\mathrm{e}^{\pi/2}$,
the gap asymptotically approaches zero in an exponential fashion;
in phase II, for $\mathrm{e}^{-\pi/2}<\Delta_{i}/\Delta_{f}<\mathrm{e}^{\pi/2}$,
the gap shows damped $t^{-1/2}$ oscillations around one asymptotic
value; and in phase III, which takes place for $\Delta_{i}/\Delta_{f}<\mathrm{e}^{-\pi/2}$,
the gap shows persistent oscillations between two asymptotic values.

Whether the two-band BCS model is integrable or not is beyond the
scope of this work. Given the difficulties in finding the integrals
of motion of the two-band case, in this section we employ instead
a perturbative method to extract the long-time asymptotic dynamics
of the superconducting gap in phase II, where the gap shows damped
oscillations. This is precisely the behavior found numerically for
weak quenches, shown in Fig. \ref{Fig_Gap_Oscillations}. In particular,
the method we develop here is a modified version of the one pioneered
by Volkov and Kogan in Ref.~\onlinecite{Volkov1974}, which allows
us to also analytically determine the steady-state gap values $\Delta_{\alpha,\infty}$.

For convenience, we briefly review our notation scheme: subscripts
$i$ and $f$ denote the thermal equilibrium value before ($i$) and
after ($f$) the quench. The subscript $\infty$ denotes the long-time
asymptotic steady-state value of the gap. For example, $\Delta_{\alpha,i}$
($\Delta_{\alpha,f}$) is the equilibrium value of gap $\alpha$ before
(after) the quench, and $\Delta_{\alpha,\infty}$ is its long-time
asymptotic steady-state value following the time evolution governed
by the BCS Hamiltonian. We note that our analysis is restricted to
weak quenches, resulting in the system being in phase II, where the
gap experiences Volkov-Kogan-like behavior.

\subsection{Linearized equations of motion}

\label{subsec:linearized_eom} To analytically describe the post-quench
gap dynamics at long times, we generalize the method used first by
Volkov and Kogan in Ref.~\onlinecite{Volkov1974}. Instead of expanding
around the final equilibrium state $S_{\alpha,f}^{i}$ and $\Delta_{\alpha,f}$,
however, we expand around the long-time non-equilibrium steady-state
values $S_{\alpha,\infty}^{i}$ and $\Delta_{\alpha,\infty}$. Importantly,
these steady-state values will be determined self-consistently in
our calculation using Laplace's final value theorem. We thus assume
that in the long-time limit the superconducting gaps reach their long-time
asymptotic values $\Delta_{\alpha,\infty}$. Specifically, we expand
the equations of motion and the gap equations around the asymptotic
steady-state values \begin{subequations} 
\begin{align}
S_{\alpha}^{z}\left(\varepsilon,t\right) & =S_{\alpha,\infty}^{z}\left(\varepsilon\right)+g_{\alpha}\left(\varepsilon,t\right)\label{Linearization_Sz}\\
S_{\alpha}^{-}\left(\varepsilon,t\right) & =S_{\alpha,\infty}^{-}\left(\varepsilon\right)+f_{\alpha}\left(\varepsilon,t\right)\\
\Delta_{\alpha}\left(t\right) & =\Delta_{\alpha,\infty}+\delta_{\alpha}\left(t\right)
\end{align}
\label{eq:linearization_around_ss_values} \end{subequations}where,
from the stationary condition of the equations of motion, $S_{\alpha,\infty}^{\pm}=S_{\alpha,\infty}^{x}$,
$S_{\alpha,\infty}^{y}=0$, $\varepsilon S_{\alpha,\infty}^{x}=-\Delta_{\alpha,\infty}S_{\alpha,\infty}^{z}$.
Note that $f_{\alpha}$ describes pairing amplitude fluctuations and
$g_{\alpha}$ describes density fluctuations. The deviation of the
gap from its long-time asymptotic value is denoted by $\delta_{\alpha}$,
which is determined by the pairing-amplitude fluctuations $f_{\alpha}$
via the gap equation:

\begin{equation}
\delta_{\alpha}\left(t\right)=v_{f}\sum_{\beta}\gamma_{\alpha\beta}\int_{-\Lambda}^{\Lambda}d\varepsilon f_{\beta}\left(\varepsilon,t\right)\label{gap_eq_redux}
\end{equation}
where $\gamma_{\alpha\beta}$ is given in Eq. \ref{eq:gamma_matrix}.\textcolor{blue}{{}
}As we will show below, because $f''_{\alpha}$ is an odd function
of $\varepsilon$, $\delta_{\alpha}$ is real, as long as we choose
the initial equilibrium gaps of the two bands $\Delta_{\alpha,i}$
to be real. With this in mind, we linearize the equations of motion
by inserting Eqs.~\eqref{eq:linearization_around_ss_values} into
Eq.~\eqref{EOM} to obtain \begin{subequations} 
\begin{align}
\dot{f}_{\alpha}' & =2\varepsilon f_{\alpha}''\\
\dot{f}_{\alpha}'' & =-2\varepsilon f_{\alpha}'-2\Delta_{\alpha,\infty}g_{\alpha}-2S_{\alpha,\infty}^{z}\delta_{\alpha}\left(t\right)\\
\dot{g}_{\alpha} & =2\Delta_{\alpha,\infty}f_{\alpha}''\label{Linearization_EOM}
\end{align}
\end{subequations}where $f_{\alpha}=f_{\alpha}'+if_{\alpha}''$ and
the notation $\dot{f}\equiv\frac{df}{dt}$ is used. Note that, as
anticipated, $f_{\alpha}''$ remains an odd function of $\varepsilon$
for all times, since $S_{z,\infty}$ and $g_{\alpha}$ are odd while
$f_{\alpha}'$ is even. As a result, the gap remains real for all
times. The fact that the phases of the gaps are constants of motion
follows directly from the particle-hole symmetry of the BCS Hamiltonian~\cite{Barankov-Synchronization}.
Therefore, the relative phase of the two gaps is also a constant of
motion and the Leggett (relative phase) mode, which would in any case
be overdamped in the regime we study here of inter-band pairing interaction
only, is not excited in our quench protocol. In order to excite it,
one must break the particle-hole symmetry of the BCS Hamiltonian,
for example, by external perturbations as in the pump-probe setups~\cite{Krull2016-Leggett-Higgs}.

The linearized equations of motion faithfully describe
the long-time dynamics since at the long-time limit, the deviations
from the asymptotic values are small, i.e. $\left(g_{\alpha},f_{\alpha},\delta_{\alpha}\right)\ll\left(S_{\alpha,\infty}^{z},S_{\alpha,\infty}^{-},\Delta_{\alpha,\infty}\right)$.
To have a better description of the gap dynamics over
a wider time range, we focus on relatively weak quenches where $v_{f}/v_{i}$
is close to 1. In this case, the oscillations around $\Delta_{\alpha,\infty}$
are small already at earlier times,
allowing for a better comparison between numerics and analytics. Such
weak quench regime is also the most relevant to experiments, where
excess heating is suppressed.

Since we are interested in $\delta_{\alpha}$, which is only related
to $f_{\alpha}$, see Eq. \ref{gap_eq_redux}, we can further simplify
the above equations by eliminating $g_{\alpha}$ to find \begin{subequations}
\begin{align}
\ddot{f}_{\alpha}'' & =-4E_{\alpha,\infty}^{2}f_{\alpha}''-2S_{\alpha,\infty}^{z}\dot{\delta}_{\alpha}\left(t\right)\label{eq:EOM_Im_f}\\
\dddot{f}_{\alpha}' & =-4E_{\alpha,\infty}^{2}\dot{f}_{\alpha}'-4\varepsilon S_{\alpha,\infty}^{z}\dot{\delta}_{\alpha}\left(t\right)\,,\label{eq:EOM_Re_f}
\end{align}
\end{subequations} where $E_{\alpha,\infty}^{2}=\varepsilon^{2}+\Delta_{\alpha,\infty}^{2}$.
Eqs. \ref{eq:EOM_Im_f} and \ref{eq:EOM_Re_f} describe the dynamics
of the imaginary and real parts of the pairing amplitude fluctuations,
respectively, which determine the time evolution of the gap.

\subsection{Solution in Laplace space}

\label{subsec:sol_laplace}

To solve the differential equations \eqref{eq:EOM_Im_f} and~\eqref{eq:EOM_Re_f},
it is useful to perform a Laplace transformation $y\left(s\right)=\int_{0}^{\infty}y\left(t\right)e^{-st}dt$.
We find the the following algebraic equations: 
\begin{widetext}
\begin{subequations} 
\begin{align}
f_{\alpha}''\left(s\right)+\frac{2sS_{\alpha,\infty}^{z}}{s^{2}+4E_{\alpha,\infty}^{2}}\delta_{\alpha}\left(s\right) & =\frac{sf_{\alpha,0}''+\dot{f}_{\alpha,0}''}{s^{2}+4E_{\alpha,\infty}^{2}}+\frac{2S_{\alpha,\infty}^{z}}{s^{2}+4E_{\alpha,\infty}^{2}}\delta_{\alpha,0}\label{EOM_Laplace_1}\\
f_{\alpha}'\left(s\right)-\frac{-4\varepsilon S_{\alpha,\infty}^{z}}{s^{2}+4E_{\alpha,\infty}^{2}}\delta_{\alpha}\left(s\right) & =\frac{1}{s}\left[f_{\alpha,0}'-\frac{-4\varepsilon S_{\alpha,\infty}^{z}}{s^{2}+4E_{\alpha,\infty}^{2}}\delta_{\alpha,0}\right]-\frac{2\varepsilon}{s}\frac{sf_{\alpha,0}''+\dot{f}_{\alpha,0}''}{s^{2}+4E_{\alpha,\infty}^{2}}\,.\label{EOM_Laplace_2}
\end{align}
\end{subequations} 
\end{widetext}

Here, $s$ is the complex frequency in the Laplace domain and the
subscript $0$ indicates an initial condition, i.e. $f_{\alpha,0}\equiv f_{\alpha}\left(\varepsilon,t=0^{+}\right)$,
$\delta_{\alpha,0}\equiv\delta_{\alpha}\left(t=0^{+}\right)$, etc.
Physically, Eqs.~\eqref{EOM_Laplace_1} and~\eqref{EOM_Laplace_2}
describe the phase and amplitude dynamics of the gap, respectively.

Since $\delta_{\alpha}$ and $f_{\alpha}$ are related through the
gap equation \ref{gap_eq_redux}, it is convenient to integrate both
sides of the above equations over $\varepsilon$. Then, Eq.~\eqref{EOM_Laplace_1}
is trivially satisfied, since $S_{\alpha,\infty}^{z}$ is an odd function
of $\varepsilon$, by virtue of Eq. \ref{aux}, $f_{\alpha,0}^{''}=0$
by construction, and $\dot{f}_{\alpha,0}''$ is an odd function of
$\varepsilon$, by virtue of the second equation of \ref{Linearization_EOM}.
Consequently, we are left with a single equation for $f_{\alpha}'$
and $\delta_{\alpha}$, which are related through the gap equation
\ref{gap_eq_redux}.

Expressing $f$ in terms of $\delta$, and recasting Eq.~\eqref{EOM_Laplace_2}
in matrix form, the deviations of the superconducting gaps from their
asymptotic values, $\delta_{\alpha}$, are given by: 
\begin{equation}
\left(\hat{\Phi}^{\infty}\left(s\right)+\hat{\mathcal{M}}\right)\vec{\delta}\left(s\right)=\frac{\vec{I}\left(s\right)}{s}\,,\label{gap_Laplace}
\end{equation}
where the hat (arrow) denote a matrix (vector) in band space. Here,
we defined: 
\begin{align}
\hat{\Phi}_{\alpha\beta}^{\infty}\left(s\right) & =\mathbb{I}_{\alpha\beta}\left(s^{2}+4\Delta_{\alpha,\infty}^{2}\right)\left\langle \frac{S_{\alpha,\infty}^{x}/\Delta_{\alpha,\infty}}{s^{2}+4E_{\alpha,\infty}^{2}}\right\rangle \\
\hat{\mathcal{M}}_{\alpha\beta} & =\left(\hat{\gamma}^{-1}\right)_{\alpha\beta}-\mathbb{I}_{\alpha\beta}\left\langle \frac{S_{\beta,\infty}^{x}}{\Delta_{\beta,\infty}}\right\rangle \,.
\end{align}
where $\mathbb{I}$ is the identity matrix in band space and the following
notation is used: 
\begin{equation}
\left\langle \ldots\right\rangle =v_{f}\int d\varepsilon\left(\ldots\right)\,.
\end{equation}

For convenience, we write $\hat{\Phi}_{\alpha\beta}^{\infty}\left(s\right)\equiv\mathbb{I}_{\alpha\beta}\Phi_{\alpha}^{\infty}$
and define:

\begin{equation}
\Phi_{\alpha}^{\infty}(s)=\left(s^{2}+4\Delta_{\alpha,\infty}^{2}\right)\left\langle \frac{S_{\alpha,\infty}^{x}/\Delta_{\alpha,\infty}}{s^{2}+4E_{\alpha,\infty}^{2}}\right\rangle \label{eq_Phi_infty}
\end{equation}

The function $\vec{I}\left(s\right)$ on the right-hand side is given
by (detailed derivation in Appendix~\ref{sec:app_initial_conditions})
\begin{align}
I_{\alpha}\left(s\right) & =\sum_{\beta}\left(\hat{\gamma}^{-1}\right)_{\alpha\beta}\delta_{\beta,0}+\left(\Delta_{\alpha,i}-\Delta_{\alpha,\infty}\right)\nonumber \\
 & \quad\times\left[\Phi_{\alpha}^{i}\left(s\right)-\frac{v_{f}}{v_{i}}\sum_{\beta}\left(\hat{\gamma}^{-1}\right)_{\alpha\beta}\frac{\Delta_{\beta,i}}{\Delta_{\alpha,i}}\right]\,,\label{Initial_Conditions}
\end{align}
with: 
\begin{equation}
\Phi_{\alpha}^{i}\left(s\right)=\left(s^{2}+4\Delta_{\alpha,\infty}^{2}\right)\left\langle \frac{S_{\alpha,i}^{x}/\Delta_{\alpha,i}}{s^{2}+4E_{\alpha,\infty}^{2}}\right\rangle \label{eq:Phi_i}
\end{equation}

The solution for $\vec{\delta}(s)$ in Laplace space is then simply
given by 
\begin{equation}
\vec{\delta}\left(s\right)=\left(\hat{\Phi}^{\infty}\left(s\right)+\hat{\mathcal{M}}\right)^{-1}\frac{\vec{I}\left(s\right)}{s}
\end{equation}

It is clear that without inter-band interaction, $V=0$, $\hat{\mathcal{M}}_{\alpha\beta}$
becomes a diagonal matrix, since $\hat{\gamma}_{\alpha\beta}$ in
Eq. \ref{eq:gamma_matrix} is diagonal. As a result, Eq. \ref{gap_Laplace}
becomes diagonal in band space as well, and the two-band model reduces
to two independent one-band models.

In the following subsections, we will extract the dynamics of the
gaps in the long-time limit from their analytic behaviors in Laplace
space. These are determined by the functions $\hat{\Phi}^{\infty}\left(s\right)$
and $\vec{I}\left(s\right)$, as they are the only $s$-dependent
functions in Eq.~\eqref{gap_Laplace}. Their $s$-dependence comes
from the two functions $\Phi_{\alpha}^{\infty}\left(s\right)$ and
$\Phi_{\alpha}^{i}\left(s\right)$ defined above.

The function $\Phi_{\alpha}^{i}\left(s\right)$ is straightforward
to calculate since the initial pseudospin configuration is given by
the equilibrium value of the gap at $T=0$, i.e. $S_{\alpha,i}^{x}/\Delta_{\alpha,i}=\frac{1}{2\sqrt{\varepsilon^{2}+\Delta_{\alpha,i}^{2}}}$.
Inserting this initial pseudo-spin state into Eq.~\eqref{eq:Phi_i},
this can be brought to the form 
\begin{equation}
\Phi_{\alpha}^{i}\left(s\right)=\Upsilon\left(\tilde{\Delta}_{\alpha,i},\frac{s}{2\Delta_{\alpha,\infty}}\right)\label{phi_initial}
\end{equation}
where we defined the dimensionless ratio $\tilde{\Delta}_{\alpha,i}=\Delta_{\alpha,i}/\Delta_{\alpha,\infty}$
and the function 
\begin{equation}
\Upsilon\left(\Delta,x\right)=v_{f}\frac{\sqrt{\frac{x^{2}+1}{\Delta^{2}}}\arccos\left(\sqrt{\frac{x^{2}+1}{\Delta^{2}}}\right)}{\sqrt{1-\frac{1+x^{2}}{\Delta^{2}}}}\label{eq:definition_Upsilon}
\end{equation}

To find an explicit expression for $\Phi_{\alpha}^{\infty}\left(s\right)$,
given by Eq. \ref{eq_Phi_infty}, we first need to compute the function
$S_{\alpha,\infty}^{x}/\Delta_{\alpha,\infty}$. The gap equation
(see Eq.~\eqref{gap_eq}), which is satisfied regardless of whether
the system is in thermal equilibrium or not, restricts the expectation
value of this quantity to: 
\begin{equation}
\left\langle \frac{S_{\alpha,\infty}^{x}}{\Delta_{\alpha,\infty}}\right\rangle =\sum_{\beta}\left(\hat{\gamma}^{-1}\right)_{\alpha\beta}\frac{\Delta_{\beta,\infty}}{\Delta_{\alpha,\infty}}
\end{equation}

As we discussed above, the non-zero inter-band interactions render
the matrix $\hat{\mathcal{M}}$ off-diagonal and make the two-band
model fundamentally different than the single-band case. While a generic
discussion of arbitrary inter- and intra-band interactions is possible,
the analysis is simplified considerably by focusing on the case of
inter-band repulsion only, i.e. $r=0$. Indeed, our numerical results
discussed in Fig. \ref{Fig_Gap_Oscillations_r} show that the general
behavior of the two-band problem is the same for $r=0$ and $r\neq0$.
Setting $r=0$ in Eq.~\eqref{eq:gamma_matrix} yields an off-diagonal
matrix $\hat{\gamma}=\left(\begin{array}{cc}
0 & -\eta\\
-1 & 0
\end{array}\right)$. As result, the equation above becomes:

\begin{equation}
\left\langle \frac{S_{1,\infty}^{x}}{\Delta_{2,\infty}}\right\rangle =\eta\left\langle \frac{S_{2,\infty}^{x}}{\Delta_{1,\infty}}\right\rangle =-1\,.\label{eq:constraint_pure_interband}
\end{equation}

Note that this ratio involves the pseudospin of band $\alpha$ and
the gap of the other band $\bar{\alpha}$, where $\bar{\alpha}=1(2)$
for $\alpha=2(1)$. To proceed, we note that, in equilibrium, the
same relationship holds between the ratios of the pseudospin and the
gap:

\begin{equation}
\left\langle \frac{S_{1,f}^{x}}{\Delta_{2,f}}\right\rangle =\eta\left\langle \frac{S_{2,f}^{x}}{\Delta_{1,f}}\right\rangle =-1\,.\label{eq:constraint_aux}
\end{equation}

The difference is that, in equilibrium, from Eq. \ref{aux}, we know
precisely the expression for $S_{\alpha,f}^{x}$:

\begin{subequations} 
\begin{align}
\left\langle \frac{S_{1,f}^{x}}{\Delta_{2,f}}\right\rangle  & =\left\langle \frac{\Delta_{1,f}/\Delta_{2,f}}{2\sqrt{\varepsilon^{2}+\Delta_{1,f}^{2}}}\right\rangle =-1\\
\eta\left\langle \frac{S_{2,f}^{x}}{\Delta_{1,f}}\right\rangle  & =\eta\left\langle \frac{\Delta_{2,f}/\Delta_{1,f}}{2\sqrt{\varepsilon^{2}+\Delta_{2,f}^{2}}}\right\rangle =-1\,,
\end{align}
\end{subequations} Based on this similarity, we propose the following
ansatz: 
\begin{equation}
\frac{S_{\alpha,\infty}^{x}}{\Delta_{\alpha,\infty}}=\frac{\tilde{\Delta}_{\alpha,f}}{\tilde{\Delta}_{\bar{\alpha},f}}\left(\frac{1}{2\sqrt{\varepsilon^{2}+\Delta_{\alpha,f}^{2}}}\right)\label{eq:ansatz}
\end{equation}
where $\tilde{\Delta}_{\alpha,f}=\Delta_{\alpha,f}/\Delta_{\alpha,\infty}$
is defined analogously to $\tilde{\Delta}_{\alpha,i}$. Clearly, this
ansatz satisfies the constraint \ref{eq:constraint_pure_interband}.
For $r\neq0$, the constraint will likely have a more complicated
form; thus, for the sake of clarity, we focus on the case $r=0.$
We will verify the validity of this ansatz later by an explicit comparison
to numerical calculations and by comparison with the exact solution
of the single-band case. For now, we proceed with this ansatz and
perform the energy integration in the expression of $\Phi_{\alpha}^{\infty}\left(s\right)$.
We obtain: 
\begin{equation}
\Phi_{\alpha}^{\infty}\left(s\right)=\frac{\tilde{\Delta}_{\alpha,f}}{\tilde{\Delta}_{\bar{\alpha},f}}\Upsilon\left(\tilde{\Delta}_{\alpha,f},\frac{s}{2\Delta_{\alpha,\infty}}\right)\label{phi_infinite}
\end{equation}

\subsection{Asymptotic gap values}

\label{subsec:gap_from_self_consistency} In this subsection, we show
how to extract the long-time asymptotic steady-state gap values $\Delta_{\alpha,\infty}$
self-consistently. To set the stage, and validate the ansatz proposed
in the previous subsection, we first present the calculation for the
single-band case, comparing the perturbative solution with the exact
one.

\subsubsection{Asymptotic gap for the single-band model}

\label{subsubsec:asymptotic_gap_1_band}In the single-band BCS model
with attractive pairing interaction $u\equiv U\mathcal{N}$, a quench
suddenly changes the pairing interaction $u_{i}\rightarrow u_{f}$.
It is convenient to use $\Delta_{i}/\Delta_{f}$ as the quench parameter,
where $\Delta_{i}$ ($\Delta_{f}$) is the equilibrium value of the
gap with pairing interaction $u_{i}$ ($u_{f}$). We employ the same
linearization scheme for the single-band model as above in Eqs.~\eqref{Linearization_Sz}-\eqref{Linearization_EOM}
for the two-band case, and expand around the long-time asymptotic
values, $S_{\infty}^{\alpha}$ and $\Delta_{\infty}$. The equation
for the gap deviation $\delta$ in Laplace space, Eq. \ref{gap_Laplace},
becomes in the single-band case: 
\begin{equation}
\delta\left(s\right)=-\left(1-\frac{u_{f}}{u_{i}}\right)\frac{\Delta_{\infty}}{s\Phi_{\infty}\left(s\right)}+\left(\Delta_{i}-\Delta_{\infty}\right)\frac{\Phi_{i}\left(s\right)}{s\Phi_{\infty}\left(s\right)}\label{eq:delta_of_s_single_band}
\end{equation}
where \begin{subequations} 
\begin{align}
\Phi_{i}\left(s\right) & =\left\langle \frac{s^{2}+4\Delta_{\infty}^{2}}{\left(s^{2}+4E_{\infty}^{2}\right)}\frac{S_{i}^{x}}{\Delta_{i}}\right\rangle \\
\Phi_{\infty}\left(s\right) & =\left\langle \frac{s^{2}+4\Delta_{\infty}^{2}}{\left(s^{2}+4E_{\infty}^{2}\right)}\frac{S_{\infty}^{x}}{\Delta_{\infty}}\right\rangle 
\end{align}
\end{subequations} Here, $S_{i}^{x}/\Delta_{i}=1/(2E_{i})$ with
$E_{i}=\sqrt{\varepsilon^{2}+\Delta_{i}^{2}}$ is given by its value
in the initial $T=0$ ground state state prior to the quench. The
ratio $S_{\infty}^{x}/\Delta_{\infty}$, according to our ansatz (\ref{eq:ansatz}),
becomes in the single-band case:

\begin{equation}
\frac{S_{\infty}^{x}}{\Delta_{\infty}}=\frac{1}{2\sqrt{\varepsilon^{2}+\Delta_{f}^{2}}}\label{single_band_ansatz}
\end{equation}

This ansatz can be recast in an alternative way as an ansatz for the
non-equilibrium distribution function. From the definition of $S_{f}^{x}$,
Eq. \ref{aux}, we have:

\begin{equation}
S_{f}^{x}=\frac{\Delta_{f}\,n_{0}(\varepsilon)}{2\sqrt{\varepsilon^{2}+\Delta_{f}^{2}}}\label{eq:single_band_Sx}
\end{equation}
where we defined the equilibrium distribution function $n_{0}(\varepsilon)=\tanh(\sqrt{\varepsilon^{2}+\Delta_{f}^{2}}/\left(2T\right)).$
From the gap equation, it follows that $\left\langle \frac{S_{f}^{x}}{\Delta_{f}}\right\rangle =1$.
Analogously, we can express $S_{\infty}^{x}$ in terms of the non-equilibrium
quasiparticle distribution function $n_{\text{eff}}\left(\varepsilon\right)$:

\begin{equation}
S_{\infty}^{x}=\frac{\Delta_{\infty}n_{\text{eff}}\left(\varepsilon\right)}{2\sqrt{\varepsilon^{2}+\Delta_{\infty}^{2}}}
\end{equation}

Because the gap equation has to be satisfied also in non-equilibrium,
it follows that: 
\begin{equation}
\left\langle \frac{S_{\infty}^{x}}{\Delta_{\infty}}\right\rangle =\left\langle \frac{n_{\text{eff}}\left(\varepsilon\right)}{2\sqrt{\varepsilon^{2}+\Delta_{\infty}^{2}}}\right\rangle =1\,.\label{eq:gap_single_band}
\end{equation}

The ansatz \ref{single_band_ansatz} thus can be recast as an ansatz
for the non-equilibrium distribution function: 
\begin{equation}
n_{\text{eff}}\left(\varepsilon\right)=n_{0}(\varepsilon)\sqrt{\frac{\varepsilon^{2}+\Delta_{\infty}^{2}}{\varepsilon^{2}+\Delta_{f}^{2}}}\,,\label{eq:n_non_eq}
\end{equation}

Having obtained an explicit expression for $S_{\infty}^{x}/\Delta_{\infty}$,
we can derive analytic expressions for $\Phi_{i}\left(s\right)$ and
$\Phi_{\infty}\left(s\right)$: 
\begin{equation}
\Phi_{i/\infty}\left(s\right)=u_{f}\frac{\sqrt{s^{2}+4\Delta_{\infty}^{2}}\arccos\left(\frac{\sqrt{s^{2}+4\Delta_{\infty}^{2}}}{2\left|\Delta_{i/f}\right|}\right)}{\sqrt{4\left(\Delta_{i/f}^{2}-\Delta_{\infty}^{2}\right)-s^{2}}}\label{eq:Phi_single_band_explicit_expression}
\end{equation}

To find the long-time asymptotic value of the gap, we use the self-consistency
condition that $\lim_{t\rightarrow\infty}\Delta\left(t\right)=\Delta_{\infty}$,
or equivalently, $\lim_{t\rightarrow\infty}\delta\left(t\right)=0$.
Using the final value theorem in Laplace space, this condition becomes
\begin{equation}
\lim_{s\rightarrow0}s\delta\left(s\right)=0\,.\label{eq:final_value}
\end{equation}

Using Eq.~\eqref{eq:delta_of_s_single_band} and inserting the explicit
expressions from Eq.~\eqref{eq:Phi_single_band_explicit_expression},
we find that the asymptotic value of the gap $\Delta_{\infty}$ must
satisfy 
\begin{equation}
\frac{\sqrt{\Delta_{f}^{2}-\Delta_{\infty}^{2}}}{\arccos\left(\frac{\Delta_{\infty}}{\Delta_{f}}\right)}\left[\ln\frac{\Delta_{i}}{\Delta_{f}}-\left(1-\frac{\Delta_{\infty}}{\Delta_{i}}\right)\frac{\arccos\left(\frac{\Delta_{\infty}}{\Delta_{i}}\right)}{\sqrt{1-\frac{\Delta_{\infty}^{2}}{\Delta_{i}^{2}}}}\right]=0\,.\label{eq:asymptotic_gap_equation}
\end{equation}

\begin{figure}[tbh]
\begin{centering}
\includegraphics[width=1\linewidth]{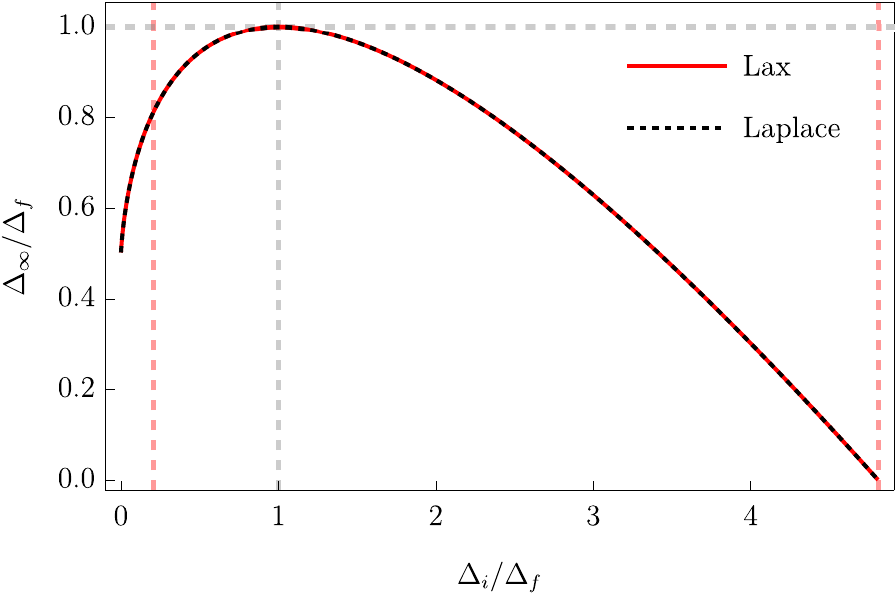} 
\par\end{centering}
\caption{Comparison of $\Delta_{\infty}$ as a function of the quench parameter
$\Delta_{i}/\Delta_{f}$ from our self-consistent perturbative method
(Eq.~\eqref{eq:asymptotic_gap_equation}) and from the exact solution
using the Lax vector technique (see Refs.~\protect\protect\onlinecite{Barankov-Synchronization,Yuzbashyan-2006-PRL}).
The vertical red dashed lines denote the extent of the phase II, as
obtained from the roots of the Lax operator. The asymptotic gap vanishes
in the phase I (larger values of $\Delta_{i}/\Delta_{f}$) and performs
persistent oscillations in the phase III (smaller values of $\Delta_{i}/\Delta_{f}$).
Our method correctly yields a vanishing gap in phase I, and provides
the average value of the gap in phase III (see Ref.~\cite{Barankov-Synchronization},
for example). \label{fig:Lax_vs_Laplace}}
\end{figure}

It is straightforward to show that this equation is identical to the
one that emerges in the exact solution of the single-band BCS gap
dynamics using the method of the Lax vector~\cite{Barankov-Synchronization,Yuzbashyan-2006-PRL}.
In Fig.~\ref{fig:Lax_vs_Laplace}, we compare the results from both
methods, which match perfectly. Interestingly, in phase III (persistent
oscillations), our method gives the average value of the gap. Of course,
our method formally breaks down in this phase, because the Laplace
final value theorem ceases to hold for an oscillatory long-time solution.

This comparison validates the ansatz \ref{eq:ansatz} for the single-band
case, giving us confidence to apply it to the two-band case as well.
Note that the perfect agreement with the exact solution does not necessarily
imply that the non-equilibrium distribution function \ref{eq:n_non_eq}
is also exact .

\subsubsection{Asymptotic gap for the two-band model}

\label{subsubsec:asymptotic_gap_2_band} We now perform the same calculation
for the two-band model with pure inter-band repulsion ($r=0$). Using
Eqs.\eqref{phi_initial} and~\eqref{phi_infinite}, we obtain the
following expression for $\delta_{\alpha}\left(s\right)$ from Eq.
\ref{gap_Laplace}: 
\begin{align}
s\delta_{\alpha}\left(s\right) & =\left(\Phi_{\bar{\alpha}}^{\infty}\left(s\right)+\frac{1}{\eta_{\bar{\alpha}}}\frac{\Delta_{\alpha,\infty}}{\Delta_{\bar{\alpha},\infty}}\right)\frac{I_{\alpha}\left(s\right)}{D\left(s\right)}+\frac{1}{\eta_{\alpha}}\frac{I_{\bar{\alpha}}\left(s\right)}{D\left(s\right)}\,,\label{sol_Laplace_1}
\end{align}
where, for convenience of notation, we introduced $\eta_{1}=1$ and
$\eta_{2}\equiv\eta$, $I_{\alpha}\left(s\right)$ is given by Eq.~\eqref{Initial_Conditions},
and: 
\begin{equation}
D\left(s\right)=\Phi_{1}^{\infty}\left(s\right)\Phi_{2}^{\infty}\left(s\right)+\frac{\Delta_{2,\infty}}{\Delta_{1,\infty}}\Phi_{2}^{\infty}\left(s\right)+\frac{1}{\eta}\frac{\Delta_{1,\infty}}{\Delta_{2,\infty}}\Phi_{1}^{\infty}\left(s\right)\,.
\end{equation}

\begin{figure}[tbh]
\begin{centering}
\includegraphics[width=1\columnwidth]{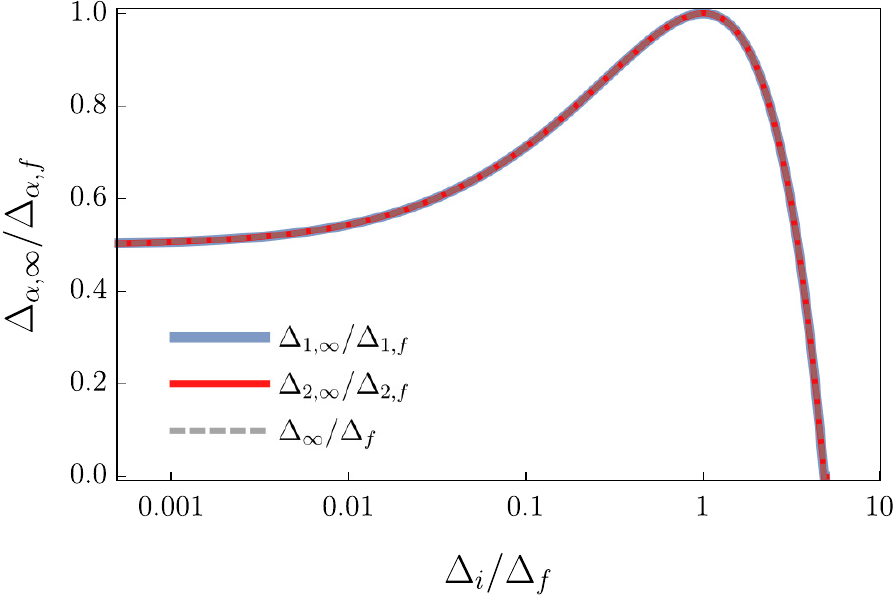} 
\par\end{centering}
\caption{Asymptotic values of the gaps in the two band case as a function of
the interaction quench parameter $\Delta_{i}/\Delta_{f}$. The dashed
gray line is the result for the single-band BCS model. For the two-band
model, we use $\Delta_{1,i}/\Delta_{1,f}$ as the quench parameter,
and we choose the ratio between the density of states to be $\eta=0.8$.
\label{fig:Del_inf}}
\end{figure}

To find the asymptotic long-time value of the gaps $\Delta_{\alpha,\infty}$,
we employ once again the final value theorem in Laplace space, Eq.
\ref{eq:final_value}. We numerically solve for $\Delta_{1,\infty}$
and $\Delta_{2,\infty}$ for a given quench protocol, $v_{i}\rightarrow v_{f}$,
or equivalently $\Delta_{1,i}\rightarrow\Delta_{1,f}$. As shown in
Fig.~\ref{fig:Del_inf}, we find that, in the case of pure inter-band
interactions ($r=0$), the ratios between the asymptotic and final
equilibrium gaps $\Delta_{\alpha,\infty}/\Delta_{\alpha,f}$ are,
to a very good approximation (i.e. with a numerical deviation of less
than $0.01\%$), equal for both bands, i.e. $\tilde{\Delta}_{1,f}=\tilde{\Delta}_{2,f}$.
They are also identical to the single-band ratio if we adjust the
definition of the quench amplitude accordingly, such that the $x$-axis
corresponds to $\Delta_{i}/\Delta_{f}$ in the single-band case and
to $\Delta_{1,i}/\Delta_{1,f}$ in the two-band case.

Using the result obtained here that $\tilde{\Delta}_{1,f}=\tilde{\Delta}_{2,f}$,
the pre-factor of Eq. \ref{phi_infinite} becomes 1. Thus, both $\Phi_{\alpha}^{\infty}\left(s\right)$
and $\Phi_{\alpha}^{i}\left(s\right)$ have the same functional dependence:
$\Phi_{\alpha}^{\infty}\left(s\right)=\Upsilon\left(\tilde{\Delta}_{\alpha,f},\frac{s}{2\Delta_{\alpha,\infty}}\right)$,
$\Phi_{\alpha}^{i}\left(s\right)=\Upsilon\left(\tilde{\Delta}_{\alpha,i},\frac{s}{2\Delta_{\alpha,\infty}}\right)$.

\subsection{Damped gap oscillations in the long-time limit}

\label{subsec:gap_oscillations} The long-time behavior of the gap
in the time domain $\Delta(t)$ can be obtained by applying the inverse
Laplace transformation to Eq.~\eqref{sol_Laplace_1}. In order to
perform the inverse Laplace transformation, we first need to study
the analytical behavior of the solution in Laplace space and find
its poles and branch cuts. They are determined by the analytic properties
of the function $\Upsilon(\Delta,x)$, defined in Eq.~\eqref{eq:definition_Upsilon}
and repeated here for convenience:

\begin{equation}
\Upsilon\left(\Delta,x\right)=v_{f}\frac{\sqrt{\frac{x^{2}+1}{\Delta^{2}}}\arccos\left(\sqrt{\frac{x^{2}+1}{\Delta^{2}}}\right)}{\sqrt{1-\frac{1+x^{2}}{\Delta^{2}}}}
\end{equation}

The reason why only the analytical properties of $\Upsilon(\Delta,x)$
matter is because we can express both $\Phi_{\alpha}^{i}(s)$ and
$\Phi_{\alpha}^{\infty}(s)$ in terms of this function: \begin{subequations}
\begin{align}
\Phi_{1}^{i/\infty}\left(s\right) & =\Upsilon\left(\tilde{\Delta}_{1,i/f},z\right)\\
\Phi_{2}^{i/\infty}\left(s\right) & =\Upsilon\left(\tilde{\Delta}_{2,i/f},\kappa z\right)\,,
\end{align}
\end{subequations} where $z=\frac{s}{2\Delta_{1,\infty}}$, $\tilde{\Delta}_{\alpha,i/f}=\frac{\Delta_{\alpha,i/f}}{\Delta_{\alpha,\infty}}$,
and $\kappa=\frac{\Delta_{1,\infty}}{\Delta_{2,\infty}}$. For concreteness,
in this section we consider the gap with $\alpha=1$ to be the one
that is asymptotically smaller, implying that $\left|\kappa\right|<1$.
But note that our results can be straightforwardly applied also to
the case $\left|\kappa\right|>1$.

The function $\Upsilon\left(\Delta,\,z\right)$ has two branch cuts,
one between $\left(-i\infty,\ -i\right)$ and another one between
$\left(i,\ i\infty\right)$. The function is analytic elsewhere. Applying
the Cauchy's residue theorem (see Appendix~\ref{sec:appendix_inverse_laplace_tf}
and Fig.~\ref{Fig_Bromwich_Contour} for details), we convert the
Bromwich integral into four integrals along the sides of the two branch
cuts. Note that we have already eliminated the pole at the origin
by imposing the final value theorem in Section~\ref{subsubsec:asymptotic_gap_1_band}.
In addition, we also use the following properties of the function
$\Upsilon$: \begin{subequations} 
\begin{align}
\Upsilon\left(\Delta,\,z\right) & =\Upsilon\left(\Delta,\,-z\right)\\
\text{Re}\left[\Upsilon\left(\Delta,\,0^{+}\pm iy\right)\right] & =\text{Re}\left[\Upsilon\left(\Delta,\,0^{-}\pm iy\right)\right],\,\text{for \ensuremath{y>1}}\\
\text{Im}\left[\Upsilon\left(\Delta,\,0^{+}\pm iy\right)\right] & =-\text{Im}\left[\Upsilon\left(\Delta,\,0^{-}\pm iy\right)\right],\,\text{for \ensuremath{y>1}.}
\end{align}
\end{subequations}

\begin{figure}[tbh]
\centering{}\centering{}\includegraphics[width=1\columnwidth]{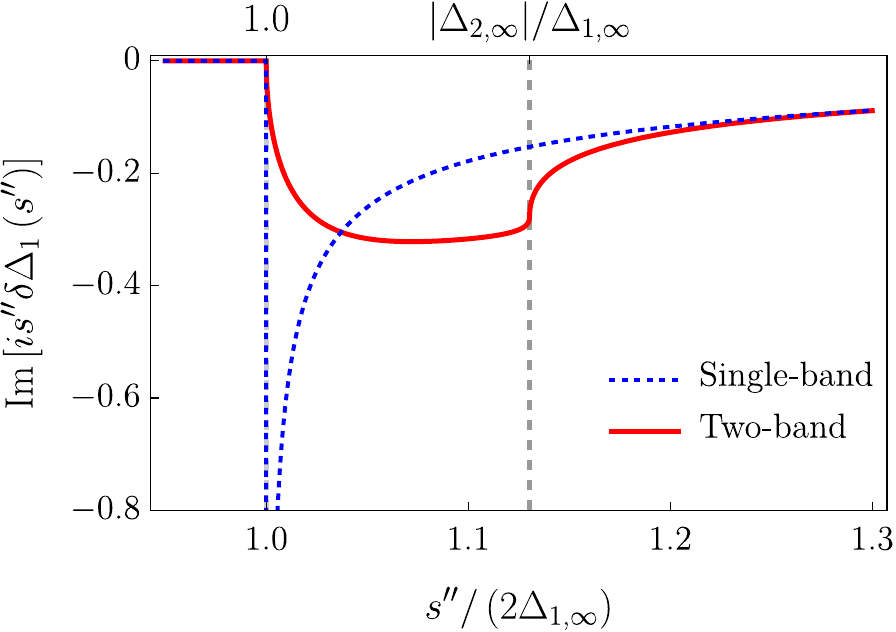}\caption{Non-analyticity of the gaps in Laplace-space along the imaginary axis,
$s''$. In the single-band case (blue dashed line), the only non-analyticity
is the inverse square root branch point at $s''=\pm2\Delta_{\infty}$
(only the positive axis is shown here). In two-band systems (red solid
line), however, the branch point at $s''=\pm2\Delta_{1,\infty}$ becomes
square root like. Moreover, additional square root branch points appear
at $s''=\pm2\Delta_{2,\infty}$, which gives rise to the additional
oscillation frequency of the gaps.}
\label{Fig_Sol-Laplace}
\end{figure}

As a result, the inverse Laplace transformation is given by the following
integral: 
\begin{equation}
\delta_{\alpha}\left(t\right)=\frac{2}{\pi}\int_{i}^{i\infty}\text{Im}\left[z\delta_{\alpha}\left(z\right)\right]\frac{\cosh\left(2\Delta_{1,\infty}zt\right)}{z}dz\label{Inverse_Laplace}
\end{equation}
where $z\delta_{\alpha}\left(z\right)$ is given by\begin{widetext}
\begin{align}
\frac{z\delta_{\alpha}\left(z\right)}{2\Delta_{\alpha,\infty}} & =-\frac{1}{\eta_{2}}\left[\frac{1}{2}\frac{v_{f}}{v_{i}}\left(\frac{\tilde{\Delta}_{\alpha,i}}{\tilde{\Delta}_{\bar{\alpha},i}}+\frac{\tilde{\Delta}_{\bar{\alpha},i}}{\tilde{\Delta}_{\alpha,i}}\right)-1\right]\frac{1}{\tilde{D}\left(z\right)}+\frac{\left(\tilde{\Delta}_{\alpha,i}-1\right)}{2}\frac{\Upsilon\left(\tilde{\Delta}_{\alpha,i},\frac{\Delta_{1,\infty}}{\Delta_{\alpha,\infty}}z\right)\Upsilon\left(\tilde{\Delta}_{\bar{\alpha},f},\frac{\Delta_{1,\infty}}{\Delta_{\bar{\alpha},\infty}}z\right)}{\tilde{D}\left(z\right)}\nonumber \\
 & \qquad-\frac{1}{2\eta_{\alpha}}\left(\frac{\Delta_{\bar{\alpha},\infty}}{\Delta_{\alpha,\infty}}\right)\left(\frac{v_{f}}{v_{i}}\frac{\tilde{\Delta}_{\bar{\alpha},i}}{\tilde{\Delta}_{\alpha,i}}-1\right)\frac{\Upsilon\left(\tilde{\Delta}_{\bar{\alpha},f},\frac{\Delta_{1,\infty}}{\Delta_{\bar{\alpha},\infty}}z\right)}{\tilde{D}\left(z\right)}\nonumber \\
 & \qquad+\frac{\left(\tilde{\Delta}_{\bar{\alpha},i}-1\right)}{2\eta_{\alpha}}\left(\frac{\Delta_{\bar{\alpha},\infty}}{\Delta_{\alpha,\infty}}\right)\frac{\Upsilon\left(\tilde{\Delta}_{\bar{\alpha},i},\frac{\Delta_{1,\infty}}{\Delta_{\bar{\alpha},\infty}}z\right)}{\tilde{D}\left(z\right)}+\frac{\left(\tilde{\Delta}_{\alpha,i}-1\right)}{2\eta_{\bar{\alpha}}}\left(\frac{\Delta_{\alpha,\infty}}{\Delta_{\bar{\alpha},\infty}}\right)\frac{\Upsilon\left(\tilde{\Delta}_{\alpha,i},\frac{\Delta_{1,\infty}}{\Delta_{\alpha,\infty}}z\right)}{\tilde{D}\left(z\right)}\label{eq:delta_alpha_Laplace}
\end{align}
with

\begin{equation}
D\left(z\right)=\Upsilon(\tilde{\Delta}_{1,f},z)\Upsilon(\tilde{\Delta}_{2,f},\kappa z)+\frac{1}{\kappa}\Upsilon(\tilde{\Delta}_{2,f},\kappa z)+\frac{\kappa}{\eta_{2}}\Upsilon(\tilde{\Delta}_{1,f},z)\,.
\end{equation}
\end{widetext}In the long-time limit, where $2\Delta_{1,\infty}t\gg1$,
the integrand of Eq.~\eqref{Inverse_Laplace} is highly oscillatory.
Only singular behaviors of $\text{Im}\left[z\delta_{\alpha}\left(z\right)\right]$
will therefore make a contribution to the long-time dynamics of the
superconducting gap. Indeed, $\text{Im}\left[z\delta_{\alpha}\left(z\right)\right]$
has two branch points along $z\in\left[i,\,i\infty\right)$: one is
located at $z=i$ and the other one is located at $z=i/\left|\kappa\right|$.
We expand $\text{Im}\left[z\delta_{\alpha}\left(z\right)\right]$
near these two branch points, i.e. $z=i+i\epsilon$ and $z=i/\left|\kappa\right|\pm i\epsilon$,
and find that both exhibit $\sqrt{\epsilon}$ behavior (details shown
in Appendix \eqref{sec:appendix_asymptotic_analysis_gap}). This is
sharply distinct from the single-band case, where only one branch
point is present along $z\in\left[i,\,i\infty\right)$. More importantly,
the asymptotic behavior in the vicinity of the branch point in the
single-band case is $1/\sqrt{\epsilon}$ rather than $\sqrt{\epsilon}$.
The two cases are plotted and compared in Fig.~\ref{Fig_Sol-Laplace}.
The $1/\sqrt{\epsilon}$ behavior leads to a $t^{-1/2}$ decay of
the gap oscillation amplitude at long times in the single-band case~\cite{Volkov1974}.
In contrast, the $\sqrt{\epsilon}$ behavior in Laplace space leads
to a faster $t^{-3/2}$ decay in the two-band model

\begin{equation}
\int_{1}^{\infty}\frac{\sqrt{y-1}}{y}\cos\left[y\left(2\Delta t\right)\right]dy\simeq-\frac{\sqrt{\pi}\sin\left(2\Delta t+\frac{\pi}{4}\right)}{2\left(2\Delta t\right)^{3/2}}
\end{equation}
for $2\Delta t\gg1$ (details are shown in Appendix~\ref{sec:appendix_inverse_laplace_tf}).
The damping of the gap oscillations thus occurs faster for two-band
superconductivity. 

To find the full long-time expressions of the gap, including prefactors
and oscillatory factors, we perform a careful asymptotic analysis
of $\text{Im}\left[z\delta_{\alpha}\left(z\right)\right]$. The final
result for the long-time gap oscillations reads

\begin{widetext}\begin{subequations} 
\begin{align}
\Delta_{1}\left(t\right) & \simeq\Delta_{1,\infty}+\mathcal{A}_{1}\frac{\sin\left(2\Delta_{1,\infty}t+\frac{\pi}{4}\right)}{\left(\Delta_{1,\infty}t\right)^{3/2}}+\mathcal{B}_{1}\frac{\sin\left(2\left|\Delta_{2,\infty}\right|t-\frac{\pi}{4}\right)}{\left(\left|\Delta_{2,\infty}\right|t\right)^{3/2}}+\mathcal{C}_{1}\frac{\sin\left(2\left|\Delta_{2,\infty}\right|t+\frac{\pi}{4}\right)}{\left(\left|\Delta_{2,\infty}\right|t\right)^{3/2}}\label{analytics_delta_1}\\
\Delta_{2}\left(t\right) & \simeq\Delta_{2,\infty}+\mathcal{A}_{2}\frac{\sin\left(2\left|\Delta_{2,\infty}\right|t+\frac{\pi}{4}\right)}{\left(\left|\Delta_{2,\infty}\right|t\right)^{3/2}}+\mathcal{B}_{2}\frac{\sin\left(2\Delta_{1,\infty}t-\frac{\pi}{4}\right)}{\left(\Delta_{1,\infty}t\right)^{3/2}}+\mathcal{C}_{2}\frac{\sin\left(2\Delta_{1,\infty}t+\frac{\pi}{4}\right)}{\left(\Delta_{1,\infty}t\right)^{3/2}}\label{analytics_delta_2}
\end{align}
\end{subequations}\end{widetext}where the pre-factors $\mathcal{A}_{\alpha}$,
$\mathcal{B}_{\alpha}$ and $\mathcal{C}_{\alpha}$ are calculated
from the asymptotic analysis and explicitly shown in Appendix~\ref{sec:appendix_asymptotic_analysis_gap}.
The gap oscillation frequencies are determined by the asymptotic values
of the gaps in the two different bands $\Delta_{\alpha,\infty}$.
As discussed in the previous sections, the asymptotic values of the
gaps are determined by the quench amplitude $\Delta_{\alpha,i}/\Delta_{\alpha,f}$
and the ratio of the density of states $\eta$ between the two bands.
In general, they will also depend on $r=-U/V$, which we have set
to zero for simplicity here. The same holds for the prefactors of
the sinusoidal oscillations.

\begin{figure*}
\begin{centering}
\includegraphics[width=0.75\paperwidth]{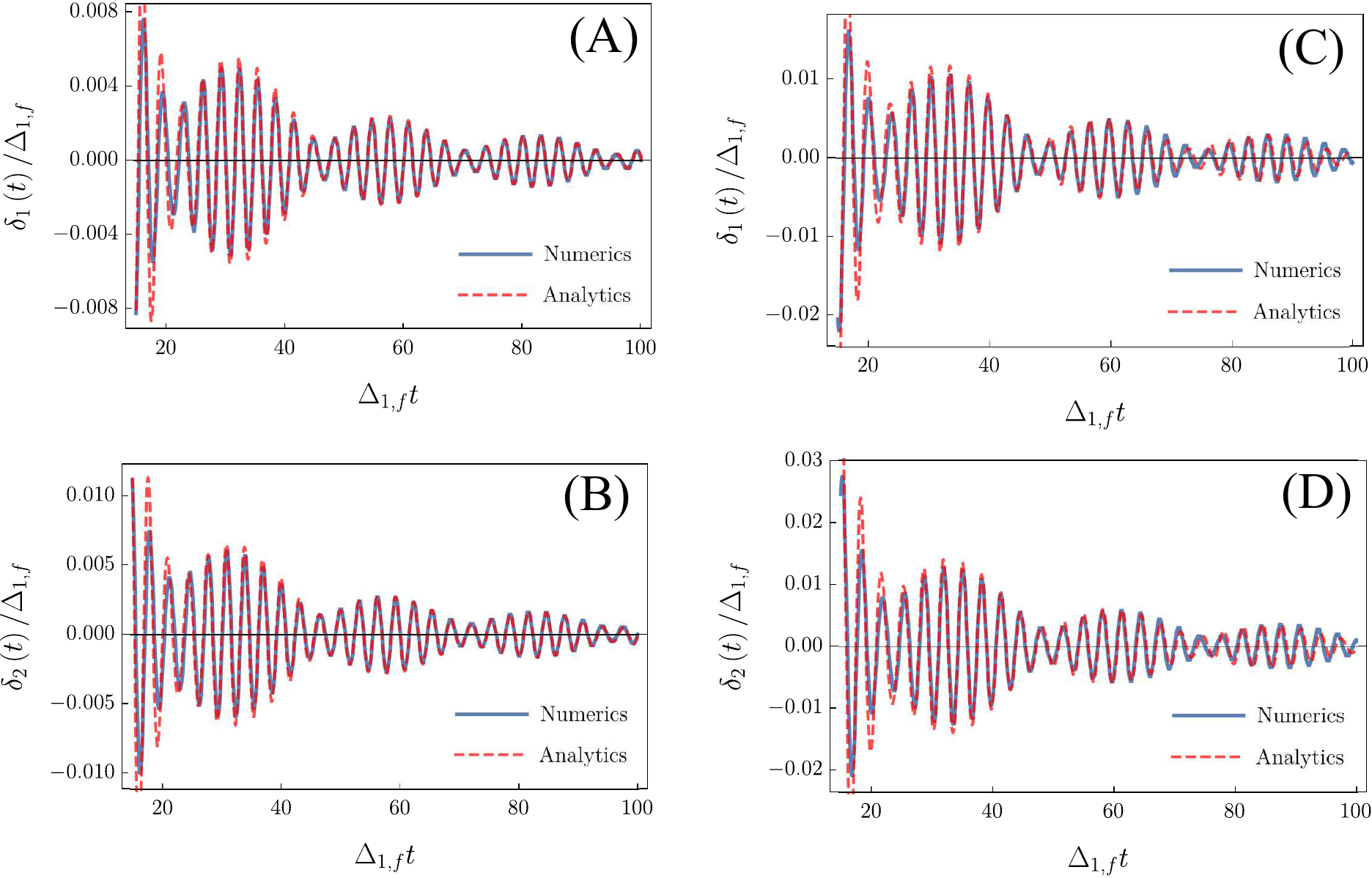}
\par\end{centering}
\caption{Comparison between the numerical solution of the gap dynamics and
the analytical approximation in Eqs.~\eqref{analytics_delta_1} and
\ref{analytics_delta_2}. (A) and (B) correspond to an interaction
quench from $v_{i}=0.19$ to $v_{f}=0.2$. (C) and (D) correspond
to an interaction quench from $v_{i}=0.18$ to $v_{f}=0.2$. The ratio
between the densities of states of the two bands is set to be $\eta=0.8$
for all panels. \label{fig:analytics_vs_numerics}}
\end{figure*}

In Fig.~\ref{fig:analytics_vs_numerics}, we compare our analytical
results to the numerical solution of the equations of motion for two
different weak quench amplitudes in phase II. We find an excellent
quantitative agreement between the two, which also justifies our analytical
ansatz \emph{a posteriori}.

We finish this section by commenting on how our solution gives the
known single-band result in the limit where the ratio between the
two densities of states approaches one, $\eta\rightarrow1$. In this
limit, the gaps have the same asymptotic magnitude, i.e. $\Delta_{1,\infty}=\left|\Delta_{2,\infty}\right|$.
The equilibrium gaps also have the same magnitude, leading to $\Upsilon\left(\tilde{\Delta}_{\alpha,f},z\right)=\Upsilon\left(\tilde{\Delta}_{\bar{\alpha},f},z\right)=\Upsilon\left(\tilde{\Delta}_{1,f},z\right)$.
As a result, Eq. \ref{eq:delta_alpha_Laplace} becomes \begin{widetext}
\begin{equation}
\frac{z\delta_{\alpha}\left(z\right)}{2\Delta_{\alpha,\infty}}=\left[\frac{1}{2}\left(\frac{v_{f}}{v_{i}}-1\right)+\frac{\left(\tilde{\Delta}_{1,i}-1\right)}{2}\Upsilon\left(\tilde{\Delta}_{1,i},z\right)\right]\frac{\left[\Upsilon\left(\tilde{\Delta}_{1,f},z\right)-2\right]}{\tilde{D}\left(z\right)}
\end{equation}
where $D\left(z\right)=\Upsilon^{2}(\tilde{\Delta}_{1,f},z)-2\Upsilon(\tilde{\Delta}_{1,f},z)$.
Further simplification of the above equation gives:
\begin{equation}
\frac{z\delta_{\alpha}\left(z\right)}{2\Delta_{\alpha,\infty}}=\frac{1}{2}\left(\frac{v_{f}}{v_{i}}-1\right)\frac{1}{\Upsilon(\tilde{\Delta}_{\alpha,f},z)}+\frac{\left(\tilde{\Delta}_{\alpha,i}-1\right)}{2}\frac{\Upsilon\left(\tilde{\Delta}_{\alpha,i},z\right)}{\Upsilon(\tilde{\Delta}_{\alpha,f},z)}
\end{equation}
\end{widetext}

In writing this last equation, we used the fact that $\Upsilon\left(\tilde{\Delta}_{\alpha,f},z\right)=\Upsilon\left(\tilde{\Delta}_{1,f},z\right)$.
This is the same expression as the solution of the single-band case
in Laplace-space, Eq. \ref{eq:delta_of_s_single_band}. Using the
asymptotic behavior of $\Upsilon\left(\tilde{\Delta}_{\alpha},iy\right)$
near the branch point $y\rightarrow1$ (details shown in Appendix
\ref{sec:appendix_asymptotic_analysis_gap}, see Eq. \ref{eq:Asymptote_Upsilon}),
we arrive at the following asymptotic behavior:
\begin{equation}
\text{Im}\left[iy\delta_{\alpha}\left(y\right)\right]\simeq\frac{v_{f}^{-1}-v_{i}^{-1}}{\pi}\left|\Delta_{\alpha,f}\right|\sqrt{\frac{2}{y-1}}
\end{equation}

By applying the inverse Laplace transformation, we find that the gap
dynamics is characterized by oscillations with frequency $2\Delta_{\infty}$
and $t^{-1/2}$ damping:
\begin{equation}
\Delta_{\alpha}\left(t\right)\simeq\Delta_{\alpha,\infty}+\left(\frac{2}{\pi}\right)^{3/2}\Delta_{\alpha,f}\ln\left(\frac{\Delta_{\alpha,i}}{\Delta_{\alpha,f}}\right)\frac{\cos\left(2\Delta_{\alpha,\infty}t+\frac{\pi}{4}\right)}{\sqrt{2\Delta_{\alpha,\infty}t}}
\end{equation}

\section{Conclusions}

\label{sec:conclusions} In this paper, we developed a generalization
of the Volkov-Kogan Laplace-space analysis for the post-quench dynamics
of $s$-wave BCS superconductors in the collisionless regime~\cite{Volkov1974},
and applied it to interaction quenches of two-band BCS superconductors.
We showed that the two-band case is fundamentally different from the
single-band case. Not only do the gap oscillations display beating
associated with the two different gap values on the two bands, but
they also display a faster $t^{-3/2}$ power-law damping, as opposed
to the $t^{-1/2}$ damping of the single-band case. For weak quenches,
our analytical results agree very well with the numerical results
in the long-time limit, demonstrating that the gap dynamics of multi-band
systems cannot be simply decomposed into the sum of the gap dynamics
of single-band systems. Formally, this new power-law decay can be
understood as arising from the ``splitting'' of the relevant branch
point in Laplace space in two, as shown in Fig. \eqref{Fig_Sol-Laplace}.
As a result, one expects the same $t^{-3/2}$ behavior to take place
even when the number of bands is larger than $2$. From a more physical
perspective, the stronger damping in the two-band case arises because
the Cooper-pairs dephasing involves states from both bands due to
the inter-band coupling. Such a dephasing is thusintrinsic to multi-band
systems and independent on the quench amplitude.

From a methodological viewpoint, our analysis is distinct from the
one introduced by Volkov and Kogan~\cite{Volkov1974} (see also the
more recent works by Yuzbashyan and co-workers~\cite{YuzbashyanDzero-PRL-2006,Yuzbashyan_Dzero_Gurarie_Foster-PRA-2015}),
because we linearize the equations of motion around the asymptotic
long-time pseudo-spin states as opposed to the final equilibrium states.
This allows us to self-consistently determine the asymptotic long-time
steady-state values of the gaps over the full range of quench amplitudes
in phase II (and phase I, where the steady-state gaps vanish). We
explicitly showed that the self-consistent equation for the steady-state
gap in the single-band case agrees with the exact expression derived
within the Lax vector analysis~\cite{Yuzbashyan_Altshuler_Enolskii-PRB-2005,Barankov-Synchronization,Yuzbashyan-2006-PRL}.
Like in the two-band model we investigate here, our method can be
very useful in cases where an exact solution is not (yet) available,
for example, to investigate quenches towards more exotic fully gapped
pairing states such as $s+is$ or $s+id$. Other interesting future
directions are to include a finite intra-band pairing interaction
$r\neq0$, competing electronic order parameters such as spin-density
waves \cite{Schuett18}, or generalize and apply our Laplace method
to study quenches in superconductors with a nodal gap structure such
as those with $d$-wave symmetry\cite{Peronaci-PRL-2015-d-wave}.
\begin{acknowledgments}
This work was supported by the by U.S. Department of Energy, Office
of Science, Basic Energy Sciences, under Award DE-SC0012336. The authors
acknowledge the Minnesota Supercomputing Institute (MSI) at the University
of Minnesota, where the numerical calculations were performed. P.P.O.
acknowledges support from Iowa State University Startup Funds. T.C.
also acknowledges the support from the Doctoral Dissertation Fellowship
awarded by the University of Minnesota. 
\end{acknowledgments}

\appendix

\section{Initial conditions for the interaction quench}

\label{sec:app_initial_conditions} The system is at equilibrium before
the interaction quench. For systems with only inter-band repulsion,
the superconducting gap is given by \begin{subequations} 
\begin{align}
\Delta_{1,i} & =-v_{i}\eta\int d\varepsilon\frac{\Delta_{2,i}}{2\sqrt{\varepsilon^{2}+\Delta_{2,i}^{2}}}\\
\Delta_{2,i} & =-v_{i}\int d\varepsilon\frac{\Delta_{1,i}}{2\sqrt{\varepsilon^{2}+\Delta_{1,i}^{2}}}
\end{align}
\end{subequations}where $v_{i}=V_{i}\mathcal{N}_{1}$ is the dimensionless
inter-band repulsion, and $\eta=\frac{\mathcal{N}_{2}}{\mathcal{N}_{1}}$
is the ratio between the density of states near the Fermi level of
the two bands. The pseudospins are \begin{subequations} 
\begin{align}
S_{\alpha,i}^{x} & =\frac{\Delta_{\alpha,i}}{2\sqrt{\varepsilon^{2}+\Delta_{\alpha,i}^{2}}}\label{initial_Sx}\\
S_{\alpha,i}^{y} & =0\label{initial_Sy}\\
S_{\alpha,i}^{z} & =\frac{-\varepsilon}{2\sqrt{\varepsilon^{2}+\Delta_{\alpha,i}^{2}}}\label{initial_Sz}
\end{align}
\end{subequations}After the interaction quench, the inter-band repulsion
is suddenly changed to a different value, $v_{f}$. The initial conditions
of the post-quench dynamics of the superconducting gaps are thus given
by replacing the inter-band repulsion with its post-quench value $v_{f}$.\begin{subequations}
\begin{align}
\Delta_{1}\left(0^{+}\right) & =-v_{f}\eta\int d\varepsilon\frac{\Delta_{2,i}}{2\sqrt{\varepsilon^{2}+\Delta_{2,i}^{2}}}=\frac{v_{f}}{v_{i}}\Delta_{1,i}\\
\Delta_{2}\left(0^{+}\right) & =-v_{f}\int d\varepsilon\frac{\Delta_{1,i}}{2\sqrt{\varepsilon^{2}+\Delta_{1,i}^{2}}}=\frac{v_{f}}{v_{i}}\Delta_{2,i}
\end{align}
\end{subequations} Substituting in the linearized equations \ref{Linearization_Sz}
and \ref{Linearization_EOM}, the initial conditions on the pseudospin
deviations $f_{\alpha}$ become \begin{subequations} 
\begin{align}
f_{\alpha,0}'' & =0\\
\dot{f}_{\alpha,0}'' & =-\frac{\varepsilon\left(\Delta_{\alpha,i}-\Delta_{\alpha,\infty}\right)}{\sqrt{\varepsilon^{2}+\Delta_{\alpha,i}^{2}}}-2\delta_{\alpha,0}S_{\alpha,\infty}^{z}
\end{align}
\end{subequations}We recall that $f_{\alpha,0}''$ and $\dot{f}_{\alpha,0}''$
are related to the dynamics of the superconducting gap in Laplace
space via $I_{\alpha}\left(s\right)=\left\langle \frac{2\varepsilon\left[sf_{\alpha,0}'+\dot{f}_{\alpha,0}''\right]}{s^{2}+4E_{\alpha,\infty}^{2}}\right\rangle $,
which yields Eq. \ref{Initial_Conditions}.

\section{Asymptotic analysis of the superconducting gap in Laplace space}
\begin{widetext}
\label{sec:appendix_asymptotic_analysis_gap}

In this appendix, we analyze the asymptotic behavior of the gap in
Laplace space near the branch points. From Eq. \ref{eq:delta_alpha_Laplace},
there are 7 terms that determine the analytic behavior of the gap.
The branch points all come from the function $\Upsilon\left(\Delta,z\right)$,
which opens branch cuts at at $\left(-i\infty,\ -i\right)$ and $\left(i,\ i\infty\right)$,
as shown in Fig. \ref{Fig_Bromwich_Contour}. Let $z=iy$, then, around
$y=1$, we have 
\begin{equation}
\Upsilon\left(\Delta,y\right)\simeq\begin{cases}
\frac{v_{f}\pi}{\left|\Delta\right|}\sqrt{\frac{1-y}{2}}+\mathcal{O}\left(1-y\right) & ,\ y\rightarrow1-\epsilon\\
i\frac{v_{f}\pi}{\left|\Delta\right|}\sqrt{\frac{y-1}{2}}+\mathcal{O}\left(y-1\right) & ,\ y\rightarrow1+\epsilon
\end{cases}\label{eq:Asymptote_Upsilon}
\end{equation}
where $\epsilon$ is an infinitesimal positive number.

We use the asymptotic behavior of $\Upsilon\left(\Delta,y\right)$
to expand all the terms in Eq. \ref{eq:delta_alpha_Laplace}, and
obtain the following results:

\begin{align*}
\text{Im}\left[\frac{1}{D\left(y\right)}\right] & \simeq\begin{cases}
-\kappa^{2}\frac{\Upsilon\left(\tilde{\Delta}_{2,f},\kappa\right)+\frac{\kappa}{\eta}}{\Upsilon^{2}\left(\tilde{\Delta}_{2,f},\kappa\right)}\frac{v_{f}\pi}{\left|\tilde{\Delta}_{1,f}\right|}\sqrt{\frac{y-1}{2}} & ,\ y\rightarrow1+\epsilon\\
\frac{\eta}{\kappa}\text{Im}\left[\frac{1}{\Upsilon\left(\tilde{\Delta}_{1,f},\frac{1}{\kappa}\right)}\right]-\left(\frac{\eta}{\kappa}\right)^{2}\text{Im}\left[\frac{\Upsilon\left(\tilde{\Delta}_{1,f},\frac{1}{\kappa}\right)+\frac{1}{\kappa}}{\Upsilon^{2}\left(\tilde{\Delta}_{1,f},\frac{1}{\kappa}\right)}\right]\frac{v_{f}\pi}{\left|\tilde{\Delta}_{1,f}\right|}\sqrt{\frac{1-\left|\kappa\right|y}{2}} & ,\ y\rightarrow\frac{1}{\left|\kappa\right|}-\epsilon\\
\frac{\eta}{\kappa}\text{Im}\left[\frac{1}{\Upsilon\left(\tilde{\Delta}_{1,f},\frac{1}{\kappa}\right)}\right]-\left(\frac{\eta}{\kappa}\right)^{2}\text{Re}\left[\frac{\Upsilon\left(\tilde{\Delta}_{1,f},\frac{1}{\kappa}\right)+\frac{1}{\kappa}}{\Upsilon^{2}\left(\tilde{\Delta}_{1,f},\frac{1}{\kappa}\right)}\right]\frac{v_{f}\pi}{\left|\tilde{\Delta}_{1,f}\right|}\sqrt{\frac{\left|\kappa\right|y-1}{2}} & ,\ y\rightarrow\frac{1}{\left|\kappa\right|}+\epsilon
\end{cases}\\
\text{Im}\left[\frac{\Upsilon\left(\tilde{\Delta}_{1,f},y\right)}{D\left(y\right)}\right] & \simeq\begin{cases}
\kappa\frac{1}{\Upsilon\left(\tilde{\Delta}_{2,f},\kappa\right)}\frac{v_{f}\pi}{\left|\tilde{\Delta}_{1,f}\right|}\sqrt{\frac{y-1}{2}} & ,\ y\rightarrow1+\epsilon\\
-\left(\frac{\eta}{\kappa}\right)^{2}\frac{1}{\kappa}\text{Im}\left[\frac{1}{\Upsilon\left(\tilde{\Delta}_{1,f},\frac{1}{\kappa}\right)}\right]\frac{v_{f}\pi}{\left|\tilde{\Delta}_{1,f}\right|}\sqrt{\frac{1-\left|\kappa\right|y}{2}} & ,\ y\rightarrow\frac{1}{\left|\kappa\right|}-\epsilon\\
-\left(\frac{\eta}{\kappa}\right)^{2}\left(1+\frac{1}{\kappa}\text{Re}\left[\frac{1}{\Upsilon\left(\tilde{\Delta}_{1,f},\frac{1}{\kappa}\right)}\right]\right)\frac{v_{f}\pi}{\left|\tilde{\Delta}_{1,f}\right|}\sqrt{\frac{\left|\kappa\right|y-1}{2}} & ,\ y\rightarrow\frac{1}{\left|\kappa\right|}+\epsilon
\end{cases}\\
\text{Im}\left[\frac{\Upsilon\left(\tilde{\Delta}_{1,i},y\right)}{D\left(y\right)}\right] & \simeq\begin{cases}
\kappa\frac{1}{\Upsilon\left(\tilde{\Delta}_{2,f},\kappa\right)}\frac{v_{f}\pi}{\left|\tilde{\Delta}_{1,i}\right|}\sqrt{\frac{y-1}{2}} & ,\ y\rightarrow1+\epsilon\\
\frac{\eta}{\kappa}\text{Im}\left[\frac{\Upsilon\left(\tilde{\Delta}_{1,i},\frac{1}{\kappa}\right)}{\Upsilon\left(\tilde{\Delta}_{1,f},\frac{1}{\kappa}\right)}\right]-\left(\frac{\eta}{\kappa}\right)^{2}\text{Im}\left[\frac{\Upsilon\left(\tilde{\Delta}_{1,i},\frac{1}{\kappa}\right)}{\Upsilon\left(\tilde{\Delta}_{1,f},\frac{1}{\kappa}\right)}+\frac{\frac{1}{\kappa}\Upsilon\left(\tilde{\Delta}_{1,i},\frac{1}{\kappa}\right)}{\Upsilon^{2}\left(\tilde{\Delta}_{1,f},\frac{1}{\kappa}\right)}\right]\frac{v_{f}\pi}{\left|\tilde{\Delta}_{1,f}\right|}\sqrt{\frac{1-\left|\kappa\right|y}{2}} & ,\ y\rightarrow\frac{1}{\left|\kappa\right|}-\epsilon\\
\frac{\eta}{\kappa}\text{Im}\left[\frac{\Upsilon\left(\tilde{\Delta}_{1,i},\frac{1}{\kappa}\right)}{\Upsilon\left(\tilde{\Delta}_{1,f},\frac{1}{\kappa}\right)}\right]-\left(\frac{\eta}{\kappa}\right)^{2}\text{Re}\left[\frac{\Upsilon\left(\tilde{\Delta}_{1,i},\frac{1}{\kappa}\right)}{\Upsilon\left(\tilde{\Delta}_{1,f},\frac{1}{\kappa}\right)}+\frac{\frac{1}{\kappa}\Upsilon\left(\tilde{\Delta}_{1,i},\frac{1}{\kappa}\right)}{\Upsilon^{2}\left(\tilde{\Delta}_{1,f},\frac{1}{\kappa}\right)}\right]\frac{v_{f}\pi}{\left|\tilde{\Delta}_{1,f}\right|}\sqrt{\frac{\left|\kappa\right|y-1}{2}} & ,\ y\rightarrow\frac{1}{\left|\kappa\right|}+\epsilon
\end{cases}\\
\text{Im}\left[\frac{\Upsilon\left(\tilde{\Delta}_{2,f},\kappa y\right)}{D\left(y\right)}\right] & \simeq\begin{cases}
\Upsilon\left(\tilde{\Delta}_{2,f},\kappa\right)\text{Im}\left[\frac{1}{D\left(y\right)}\right] & ,\ y\rightarrow1+\epsilon\\
\frac{\eta}{\kappa}\text{Im}\left[\frac{1}{\Upsilon\left(\tilde{\Delta}_{1,f},\frac{1}{\kappa}\right)}\right]\frac{v_{f}\pi}{\left|\tilde{\Delta}_{1,f}\right|}\sqrt{\frac{1-\left|\kappa\right|y}{2}} & ,\ y\rightarrow\frac{1}{\left|\kappa\right|}-\epsilon\\
\frac{\eta}{\kappa}\text{Re}\left[\frac{1}{\Upsilon\left(\tilde{\Delta}_{1,f},\frac{1}{\kappa}\right)}\right]\frac{v_{f}\pi}{\left|\tilde{\Delta}_{1,f}\right|}\sqrt{\frac{\left|\kappa\right|y-1}{2}} & ,\ y\rightarrow\frac{1}{\left|\kappa\right|}+\epsilon
\end{cases}\\
\text{Im}\left[\frac{\Upsilon\left(\tilde{\Delta}_{2,i},\kappa y\right)}{D\left(y\right)}\right] & \simeq\begin{cases}
\Upsilon\left(\tilde{\Delta}_{2,i},\kappa\right)\text{Im}\left[\frac{1}{D\left(y\right)}\right] & ,\ y\rightarrow1+\epsilon\\
\frac{\eta}{\kappa}\text{Im}\left[\frac{1}{\Upsilon\left(\tilde{\Delta}_{1,f},\frac{1}{\kappa}\right)}\right]\frac{v_{f}\pi}{\left|\tilde{\Delta}_{2,i}\right|}\sqrt{\frac{1-\left|\kappa\right|y}{2}} & ,\ y\rightarrow\frac{1}{\left|\kappa\right|}-\epsilon\\
\frac{\eta}{\kappa}\text{Re}\left[\frac{1}{\Upsilon\left(\tilde{\Delta}_{1,f},\frac{1}{\kappa}\right)}\right]\frac{v_{f}\pi}{\left|\tilde{\Delta}_{2,i}\right|}\sqrt{\frac{\left|\kappa\right|y-1}{2}} & ,\ y\rightarrow\frac{1}{\left|\kappa\right|}+\epsilon
\end{cases}\\
\text{Im}\left[\frac{\Upsilon\left(\tilde{\Delta}_{1,f},y\right)\Upsilon\left(\tilde{\Delta}_{2,i},\kappa y\right)}{D\left(y\right)}\right] & \simeq\begin{cases}
\kappa\frac{\Upsilon\left(\tilde{\Delta}_{2,i},\kappa\right)}{\Upsilon\left(\tilde{\Delta}_{2,f},\kappa\right)}\frac{v_{f}\pi}{\left|\tilde{\Delta}_{1,f}\right|}\sqrt{\frac{y-1}{2}} & ,\ y\rightarrow1+\epsilon\\
\mathcal{O}\left(\epsilon\right) & ,\ y\rightarrow\frac{1}{\left|\kappa\right|}-\epsilon\\
\frac{\eta}{\kappa}\frac{v_{f}\pi}{\left|\tilde{\Delta}_{2,i}\right|}\sqrt{\frac{\left|\kappa\right|y-1}{2}} & ,\ y\rightarrow\frac{1}{\left|\kappa\right|}+\epsilon
\end{cases}\\
\text{Im}\left[\frac{\Upsilon\left(\tilde{\Delta}_{2,f},\kappa y\right)\Upsilon\left(\tilde{\Delta}_{1,i},y\right)}{D\left(y\right)}\right] & \simeq\begin{cases}
\kappa\frac{v_{f}\pi}{\left|\tilde{\Delta}_{1,i}\right|}\sqrt{\frac{y-1}{2}} & ,\ y\rightarrow1+\epsilon\\
\frac{\eta}{\kappa}\text{Im}\left[\frac{\Upsilon\left(\tilde{\Delta}_{1,i},\frac{1}{\kappa}\right)}{\Upsilon\left(\tilde{\Delta}_{1,f},\frac{1}{\kappa}\right)}\right]\frac{v_{f}\pi}{\left|\tilde{\Delta}_{1,f}\right|}\sqrt{\frac{1-\left|\kappa\right|y}{2}} & ,\ y\rightarrow\frac{1}{\left|\kappa\right|}-\epsilon\\
\frac{\eta}{\kappa}\text{Re}\left[\frac{\Upsilon\left(\tilde{\Delta}_{1,i},\frac{1}{\kappa}\right)}{\Upsilon\left(\tilde{\Delta}_{1,f},\frac{1}{\kappa}\right)}\right]\frac{v_{f}\pi}{\left|\tilde{\Delta}_{1,f}\right|}\sqrt{\frac{\left|\kappa\right|y-1}{2}} & ,\ y\rightarrow\frac{1}{\left|\kappa\right|}+\epsilon
\end{cases}
\end{align*}

\section{Inverse Laplace transformation and useful integrals}

\label{sec:appendix_inverse_laplace_tf}

The inverse Laplace transformation is given by the Bromwich integral:
\begin{equation}
y\left(t\right)=\mathcal{L}^{-1}\left\{ Y\right\} \left(t\right)=\frac{1}{2\pi i}\int_{\sigma-i\infty}^{\sigma+i\infty}Y\left(s\right)e^{st}ds
\end{equation}
where $\sigma$ is a real number that is larger than the real parts
of all the singularities of $Y\left(s\right)$.

\begin{figure}[tbh]
\begin{centering}
\includegraphics[width=0.6\columnwidth]{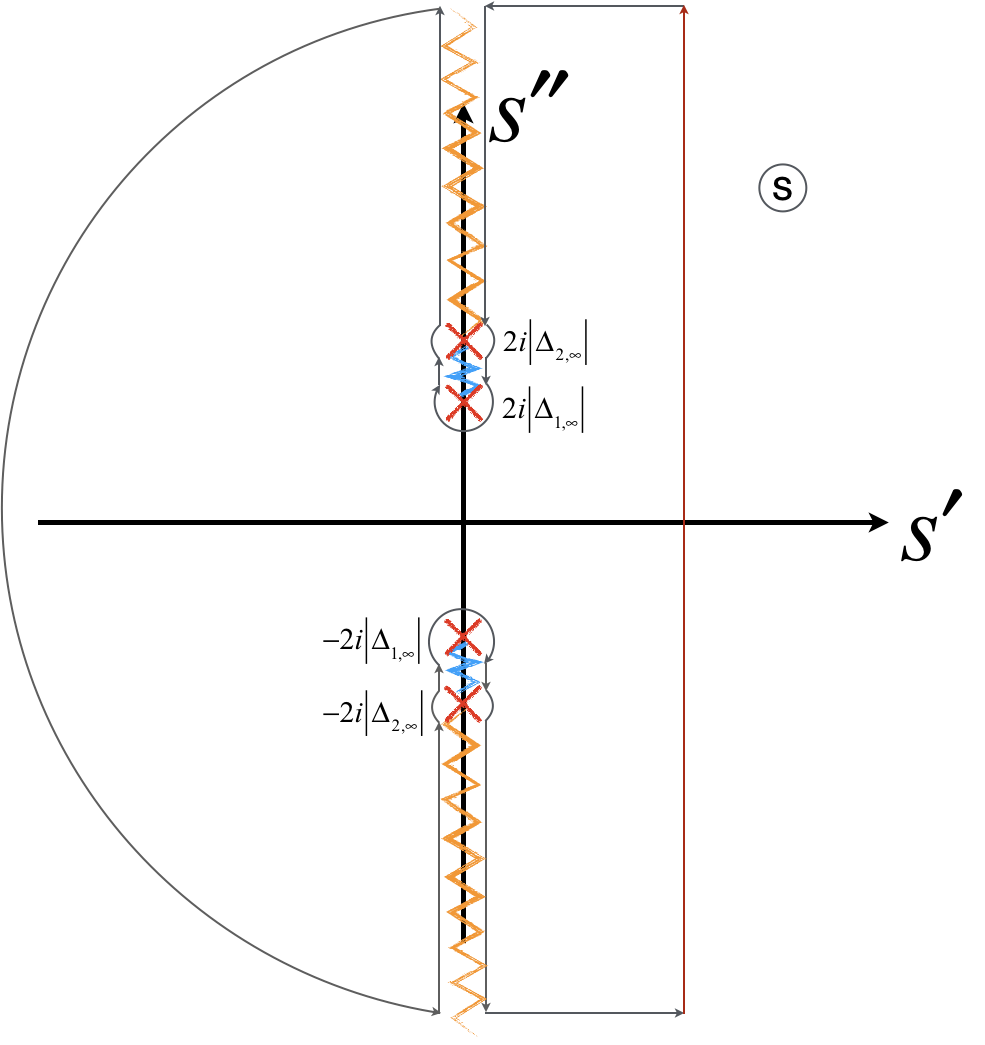} 
\par\end{centering}
\caption{Integration contour in the complex Laplace space.}
\label{Fig_Bromwich_Contour} 
\end{figure}

All the asymptotic behaviors of the gap in Laplace space are square
root like. Consequently, transforming back to real time domain leads
to a $t^{-3/2}$ decay. 
\begin{align}
\int_{1}^{\infty}\frac{\sqrt{y-1}}{y}\cos\left(2\Delta_{1,\infty}yt\right)dy & =\sqrt{\frac{\pi}{4\Delta_{1,\infty}t}}\left[\cos\left(2\Delta_{1,\infty}t\right)-\sin\left(2\Delta_{1,\infty}t\right)\right]+\pi\left[C\left(\sqrt{\frac{4\Delta_{1,\infty}t}{\pi}}\right)+S\left(\sqrt{\frac{4\Delta_{1,\infty}t}{\pi}}\right)-1\right]\nonumber \\
 & \simeq-\frac{\sqrt{\pi}\sin\left(2\Delta_{1,\infty}t+\frac{\pi}{4}\right)}{2\left(2\Delta_{1,\infty}t\right)^{3/2}}
\end{align}
for $2\Delta_{1,\infty}t\gg1$. Similarly, $\int_{\frac{1}{\left|\kappa\right|}}^{\infty}\frac{\sqrt{\left|\kappa\right|y-1}}{y}\cos\left(2\Delta_{1,\infty}yt\right)dy\simeq-\frac{\sqrt{\pi}\sin\left(2\left|\Delta_{2,\infty}\right|t+\frac{\pi}{4}\right)}{2\left(2\left|\Delta_{2,\infty}\right|t\right)^{3/2}}$,
and 
\begin{align}
\int_{-\infty}^{\frac{1}{\left|\kappa\right|}}\frac{\sqrt{1-\left|\kappa\right|y}}{y}\cos\left(2\Delta_{1,\infty}yt\right)dy & \simeq\lim_{\Lambda\rightarrow\infty}\int_{0}^{\Lambda}\sqrt{x}\cos\left(2\left|\Delta_{2,\infty}\right|t-2\left|\Delta_{2,\infty}\right|xt\right)\nonumber \\
 & \simeq\frac{\sqrt{\pi}\sin\left(2\left|\Delta_{2,\infty}\right|t-\frac{\pi}{4}\right)}{2\left(2\left|\Delta_{2,\infty}\right|t\right)^{3/2}}
\end{align}

\section{Analytical expressions for the gap dynamics}

\label{sec:appendix_analytic_expression_for_weak_quenches}

We wrote the long-time asymptotic expressions of the gap oscillations
in Eqs. \ref{analytics_delta_1} and \ref{analytics_delta_2}. In
this appendix, we provide the explicit expressions for the prefactors
$\mathcal{A}_{\alpha}$, $\mathcal{B}_{\alpha}$ and $\mathcal{C}_{\alpha}$
that appear in the two equations. 
\begin{eqnarray}
\frac{\mathcal{A}_{1}}{\Delta_{1,\infty}} & = & -\frac{\sqrt{\pi}}{4}\left\{ \left[\frac{\kappa}{\eta}\left[\frac{v_{f}}{v_{i}}\left(\frac{\tilde{\Delta}_{1,i}}{\tilde{\Delta}_{2,i}}+\frac{\tilde{\Delta}_{2,i}}{\tilde{\Delta}_{1,i}}\right)-2\right]+\left(\frac{v_{f}}{v_{i}}\frac{\tilde{\Delta}_{2,i}}{\tilde{\Delta}_{1,i}}-1\right)\Upsilon\left(\tilde{\Delta}_{2,f},\kappa\right)-\left(\tilde{\Delta}_{2,i}-1\right)\Upsilon\left(\tilde{\Delta}_{2,i},\kappa\right)\right]\times\right.\nonumber \\
 &  & \left.\times\frac{\Upsilon\left(\tilde{\Delta}_{2,f},\kappa\right)+\frac{\kappa}{\eta}}{\Upsilon^{2}\left(\tilde{\Delta}_{2,f},\kappa\right)}\frac{\kappa v_{f}}{\left|\tilde{\Delta}_{2,f}\right|}+\left(1+\frac{\kappa}{\eta}\frac{1}{\Upsilon\left(\tilde{\Delta}_{2,f},\kappa\right)}\right)\left(\tilde{\Delta}_{1,i}-1\right)\frac{\kappa v_{f}}{\left|\tilde{\Delta}_{1,i}\right|}\right\} \\
\frac{\mathcal{B}_{1}}{\Delta_{1,\infty}} & = & \frac{\sqrt{\pi}}{4}\left\{ \left(\frac{v_{f}}{v_{i}}\frac{\Delta_{1,i}}{\Delta_{2,i}}-1+\tilde{\Delta}_{2,f}-\frac{\tilde{\Delta}_{2,f}}{\tilde{\Delta}_{2,i}}\right)\text{Im}\left[\frac{1}{\Upsilon\left(\tilde{\Delta}_{1,f},\frac{1}{\kappa}\right)}\right]+\left[\frac{v_{f}}{v_{i}}\left(\frac{\tilde{\Delta}_{1,i}}{\tilde{\Delta}_{2,i}}+\frac{\tilde{\Delta}_{2,i}}{\tilde{\Delta}_{1,i}}\right)-2\right]\frac{1}{\kappa}\text{Im}\left[\frac{1}{\Upsilon^{2}\left(\tilde{\Delta}_{1,f},\frac{1}{\kappa}\right)}\right]\right.\nonumber \\
 &  & \left.-\left(\tilde{\Delta}_{1,i}-1\right)\text{Im}\left[\frac{\Upsilon\left(\tilde{\Delta}_{1,i},\frac{1}{\kappa}\right)}{\Upsilon^{2}\left(\tilde{\Delta}_{1,f},\frac{1}{\kappa}\right)}\right]\right\} \frac{\eta}{\kappa^{2}}\frac{v_{f}}{\left|\tilde{\Delta}_{2,f}\right|}\\
\frac{\mathcal{C}_{1}}{\Delta_{1,\infty}} & = & -\frac{\sqrt{\pi}}{4}\left\{ \left(\frac{v_{f}}{v_{i}}\frac{\Delta_{1,i}}{\Delta_{2,i}}-1+\tilde{\Delta}_{1,f}-\frac{\tilde{\Delta}_{1,f}}{\tilde{\Delta}_{2,i}}\right)\text{Re}\left[\frac{1}{\Upsilon\left(\tilde{\Delta}_{1,f},\frac{1}{\kappa}\right)}\right]+\left[\frac{v_{f}}{v_{i}}\left(\frac{\tilde{\Delta}_{1,i}}{\tilde{\Delta}_{2,i}}+\frac{\tilde{\Delta}_{2,i}}{\tilde{\Delta}_{1,i}}\right)-2\right]\frac{1}{\kappa}\text{Re}\left[\frac{1}{\Upsilon^{2}\left(\tilde{\Delta}_{1,f},\frac{1}{\kappa}\right)}\right]\right.\nonumber \\
 &  & \left.-\left(\tilde{\Delta}_{1,i}-1\right)\text{Re}\left[\frac{\Upsilon\left(\tilde{\Delta}_{1,i},\frac{1}{\kappa}\right)}{\Upsilon^{2}\left(\tilde{\Delta}_{1,f},\frac{1}{\kappa}\right)}\right]\right\} \frac{\eta}{\kappa^{2}}\frac{v_{f}}{\left|\tilde{\Delta}_{2,f}\right|}
\end{eqnarray}
\begin{eqnarray}
\frac{\mathcal{A}_{2}}{\Delta_{2,\infty}} & = & -\frac{\sqrt{\pi}}{4}\left\{ \left(\frac{v_{f}}{v_{i}}\frac{\Delta_{1,i}}{\Delta_{2,i}}-1+\tilde{\Delta}_{1,f}-\frac{\tilde{\Delta}_{1,f}}{\tilde{\Delta}_{1,i}}\right)\frac{1}{\Upsilon\left(\tilde{\Delta}_{2,f},\kappa\right)}+\frac{\kappa}{\eta}\left[\frac{v_{f}}{v_{i}}\left(\frac{\tilde{\Delta}_{1,i}}{\tilde{\Delta}_{2,i}}+\frac{\tilde{\Delta}_{2,i}}{\tilde{\Delta}_{1,i}}\right)-2\right]\frac{1}{\Upsilon^{2}\left(\tilde{\Delta}_{2,f},\kappa\right)}\right.\nonumber \\
 &  & \left.-\left(\tilde{\Delta}_{2,i}-1\right)\frac{\Upsilon\left(\tilde{\Delta}_{2,i},\kappa\right)}{\Upsilon^{2}\left(\tilde{\Delta}_{2,f},\kappa\right)}\right\} \frac{\kappa^{2}}{\eta}\frac{v_{f}}{\left|\tilde{\Delta}_{1,f}\right|}\\
\frac{\mathcal{B}_{2}}{\Delta_{2,\infty}} & = & \frac{\sqrt{\pi}}{4}\left\{ \left[\frac{v_{f}}{v_{i}}\left(2\frac{\tilde{\Delta}_{1,i}}{\tilde{\Delta}_{2,i}}+\frac{\tilde{\Delta}_{2,i}}{\tilde{\Delta}_{1,i}}\right)-3+\tilde{\Delta}_{2,f}-\frac{\tilde{\Delta}_{2,f}}{\tilde{\Delta}_{2,i}}\right]\text{Im}\left[\frac{1}{\Upsilon\left(\tilde{\Delta}_{1,f},\frac{1}{\kappa}\right)}\right]\right.\nonumber \\
 &  & +\left[\frac{v_{f}}{v_{i}}\left(\frac{\tilde{\Delta}_{1,i}}{\tilde{\Delta}_{2,i}}+\frac{\tilde{\Delta}_{2,i}}{\tilde{\Delta}_{1,i}}\right)-2\right]\frac{1}{\kappa}\text{Im}\left[\frac{1}{\Upsilon^{2}\left(\tilde{\Delta}_{1,f},\frac{1}{\kappa}\right)}\right]\nonumber \\
 &  & \left.-\left(\tilde{\Delta}_{1,i}-1\right)\left(\kappa\text{Im}\left[\frac{\Upsilon\left(\tilde{\Delta}_{1,i},\frac{1}{\kappa}\right)}{\Upsilon\left(\tilde{\Delta}_{1,f},\frac{1}{\kappa}\right)}\right]+\text{Im}\left[\frac{\Upsilon\left(\tilde{\Delta}_{1,i},\frac{1}{\kappa}\right)}{\Upsilon^{2}\left(\tilde{\Delta}_{1,f},\frac{1}{\kappa}\right)}\right]\right)\right\} \frac{\eta}{\kappa^{2}}\frac{v_{f}}{\left|\tilde{\Delta}_{2,f}\right|}\\
\frac{\mathcal{C}_{2}}{\Delta_{2,\infty}} & = & -\frac{\sqrt{\pi}}{4}\left\{ \left[\frac{v_{f}}{v_{i}}\left(\frac{\tilde{\Delta}_{1,i}}{\tilde{\Delta}_{2,i}}+\frac{\tilde{\Delta}_{2,i}}{\tilde{\Delta}_{1,i}}\right)-2\right]\text{Re}\left[\frac{\Upsilon\left(\tilde{\Delta}_{1,f},\frac{1}{\kappa}\right)+\frac{1}{\kappa}}{\Upsilon^{2}\left(\tilde{\Delta}_{1,f},\frac{1}{\kappa}\right)}\right]\right.\nonumber \\
 &  & +\left[\frac{v_{f}}{v_{i}}\left(\frac{\tilde{\Delta}_{1,i}}{\tilde{\Delta}_{2,i}}+\frac{\tilde{\Delta}_{2,i}}{\tilde{\Delta}_{1,i}}\right)-2+\tilde{\Delta}_{2,f}-\frac{\tilde{\Delta}_{2,f}}{\tilde{\Delta}_{2,i}}\right]\left(\kappa+\text{Re}\left[\frac{1}{\Upsilon\left(\tilde{\Delta}_{1,f},\frac{1}{\kappa}\right)}\right]\right)\nonumber \\
 &  & \left.-\left(\tilde{\Delta}_{1,i}-1\right)\text{Re}\left[\frac{\kappa\Upsilon\left(\tilde{\Delta}_{1,i},\frac{1}{\kappa}\right)}{\Upsilon\left(\tilde{\Delta}_{1,f},\frac{1}{\kappa}\right)}+\frac{\Upsilon\left(\tilde{\Delta}_{1,i},\frac{1}{\kappa}\right)}{\Upsilon^{2}\left(\tilde{\Delta}_{1,f},\frac{1}{\kappa}\right)}\right]\right\} \frac{\eta}{\kappa^{2}}\frac{v_{f}}{\left|\tilde{\Delta}_{2,f}\right|}
\end{eqnarray}
\end{widetext}

\bibliographystyle{apsrev4-1}
\bibliography{Biblio}

\end{document}